\documentclass[journal=psy]{CUP-JNL-DTM}%

\usepackage{graphicx}
\usepackage{multicol,multirow}
\usepackage{amsmath,amssymb,amsfonts}
\usepackage{mathrsfs}
\usepackage{amsthm}
\usepackage{rotating}
\usepackage{appendix}
\usepackage{ifpdf}
\usepackage[T1]{fontenc}
\usepackage{newtxtext}
\usepackage{newtxmath}
\usepackage{textcomp}
\usepackage{xcolor}
\usepackage[colorlinks,allcolors=blue]{hyperref}
\usepackage{bm}
\usepackage{algorithm}
\usepackage{algorithmic}
\usepackage{tikz}
\usetikzlibrary{arrows.meta,positioning,calc}

\theoremstyle{definition}

\numberwithin{equation}{section}

\jname{arXiv preprint}
\articletype{Article}
\jyear{2026}

\makeatletter
\def\ps@titlepage{%
     \def\@oddhead{{\hbox to \textwidth{%
      \vbox to 0pt{\vskip-7pt\rlap{\parbox{\textwidth}{\opheaderfont\begin{tabular*}{\textwidth}[b]{@{\extracolsep{\fill}}@{}l@{}}%
      \textit{\@jname}\\
      \end{tabular*}}}}%
      }}}%
     \let\@evenhead\@oddhead%
     \let\@evenfoot\@oddfoot}%
\makeatother

\begin{document}

\begin{Frontmatter}

\title[Article Title]{Frequentist forecasting in regime-switching models with extended Hamilton filter}

\author[1]{Kento Okuyama}
\author[1,2]{Tim Fabian Schaffland}
\author[1,3]{Pascal Kilian}
\author[1]{Holger Brandt}
\author[1]{Augustin Kelava}

\address[1]{\orgdiv{Methods Center}, \orgname{University of Tübingen}, \orgaddress{\city{Tübingen}, \postcode{72076}, \country{Germany}}}

\address[2]{\orgname{d-fine GmbH}, \orgaddress{\city{Frankfurt}, \postcode{60313}, \country{Germany}}}

\address[3]{\orgname{Robert Bosch GmbH}, \orgaddress{\city{Gerlingen-Schillerhöhe}, \postcode{70839}, \country{Germany}}. \\ \textbf{Corresponding author}: Kento Okuyama; \email{kento.okuyama@uni-tuebingen.de}} 

\authormark{Kento Okuyama et al.}



\abstract{Psychological change processes, such as university student dropout in math, often exhibit discrete latent state transitions and can be studied using regime-switching models with intensive longitudinal data (ILD). Recently, regime-switching state-space (RSSS) models have been extended to allow for latent variables and their autoregressive effects. Despite this progress, there is a need for frequentist estimation methods in dynamic latent variable frameworks that handle both intra-individual changes and inter-individual differences as predictors of regime-switches and allow real-time inferences and forecasts during ongoing data collection. Building on Chow and Zhang's (2013) extended Kim filter, we introduce a first frequentist filter for RSSS models which allows hidden Markov(-switching) models to depend on both latent within- and between-individual characteristics. As a counterpart of Kelava et al.'s (2022) Bayesian forecasting filter for nonlinear dynamic latent class structural equation models (NDLC-SEM), our method is the first frequentist approach within this general class of models. In an empirical study, the filter is applied to forecast emotions and behavior related to student dropout in math. A simulation study shows that marginalizing person-specific random intercepts within the filter cleanly separates autoregressive persistence from stable between-person differences, with accurate out-of-sample regime forecasts, while within-person transition effects stay weakly identified.

}

\end{Frontmatter}

\section{Introduction}
Recently, the increasing availability of intensive longitudinal data (ILD) has been brought about by digital devices such as mobile phones and smartwatches. ILD typically consists of data from several subjects, with around thirty to a hundred repeated measures per subject. The data could take various forms, such as self-report, physiological data, and behavioral, environmental, and geographic data \citep{trull2014}. Parts of such data may be time-invariant, i.e., stable across time points (\textit{inter-individual}; e.g., dispositions, personal traits). Other parts of the data consist of repeated measures whose states are subject to dynamic changes across time within individuals (\textit{intra-individual}; e.g., affective states).

Structural equation model (SEM) has been one of the most important methods in psychometrics. Dynamic SEM \citep[DSEM;][]{Asparouhov2018}, which combines SEM and time-series analysis, has been actively studied to investigate the change process in ILD. With increasing number of substantive applications of DSEMs, a number of new ILD methods have been developed \citep[e.g.,][]{Kelava2019}.

Many psychological change processes are characterized by \textit{regime-switching}, in which a change process is described by discontinuous transitions from one discrete latent state (i.e., regime) to another at a particular time point \citep[see e.g.,][]{Hamaker2012,Asparouhov2017}. Individuals are allowed to switch between regimes over time. In some cases, it is expected that the transition mostly occurs in a unidirectional manner e.g., from an intention to stay at math lecture to an intention to quit \citep[cf., Rubicon model;][]{Achtziger2018}. In other cases, subjects could repeatedly switch back and forth between two or more states such as the sleep-wake cycle \citep{Ji2020}. Studying such processes on a within-individual level requires a regime-switching model that assumes that each observation comes from one of the distinct distributions, each of which has a unique set of parameters (e.g., mean, variance, autocorrelation).

Since Markov-switching autoregressive (MSAR) models \citep{hamilton1989} were introduced, regime-switching models have mostly gained the interest of econometricians. \cite{KIM1994} generalized this approach to state-space models, which are called regime-switching state-space (RSSS) models. This framework accommodated a wide range of models, including latent variable models and regime-dependent time-series models \citep{Yang2010}.

\cite{Chow2013} then developed nonlinear RSSS models, whose state-space representation is governed by nonlinear dynamics in the state equations (handled via an extended Kalman filter) in the autoregressive time-series model. The transition between regimes is characterized by a set of transition probabilities that are invariant across individuals over time. Their estimation approach, called the extended Kim filter, combines the extended Kalman filter and the Hamilton filter to handle the estimation of parameters and latent variables in their nonlinear RSSS models. Filtering \citep{box2015} is an important technique as it allows forecasting i.e., real-time inference of unrealized future states and/or events based on available information. Upon each measurement, the prediction is continuously updated, making it suitable for applications which involve real-time monitoring and timely intervention. Upon arrival of the entire time-series observation, their estimation of the latent variables and the regime probabilities could be refined using a smoothing procedure called the extended Kim smoother.

\cite{Kelava2022} addressed forecasting the precursor of student dropout, referred to as the \textit{intention to dropout}, within their regime-switching modeling framework called the nonlinear dynamics latent class structural equation models (NDLC-SEM). Their flexible NDLC-SEM notation accommodates interactions between \textit{intra-individual} factors and \textit{inter-individual} factors in both their regime-switching model and their state-space model. An estimation procedure called the Forward Filtering Backward Sampling \citep[FFBS;][]{West1997} is implemented as a Bayesian counterpart to the extended Kim filter and extended Kim smoother. Accounting for unobserved heterogeneity in the hidden Markov(-switching) models characterizes the novelty of this approach.

\subsection*{Limitations and scope}
These methods have enabled the estimation of diverse nonlinear RSSS models. Notably, they allowed for real-time inferences and forecasts of latent or observed variables based on ongoing collections of ILD. However, there are three limitations in the current state of RSSS models.
First, the NDLC-SEM provides a very flexible approach of modeling RSSS models (including intra-individual factors for the prediction of regime membership). It is implemented using a Bayesian estimator which may become computationally very expensive, as it tries to obtain the complete parameter distribution. On the other hand, the goal of the frequentist estimator is to find a point estimate where the likelihood is maximized. Second, existing frequentist approaches do not allow for any forms of intra-individual change, inter-individual differences, and their interaction in transition probabilities. This heavily impacts their use for dynamic modeling of transition probabilities. From an applied perspective, a frequentist estimator that can include dynamic information in updating the state membership is an important next step. Third, the specification of priors is not necessary in the frequentist domain. These models are very complex and theories very often have very vague specification of expected distributions (if at all). Therefore, the lack of theoretical and empirical anticipation of effects (priors/beliefs) could be a motivation to adapt a frequentist approach.

In this paper, we present a frequentist estimation approach for RSSS models which considers both intra- and inter-individual differences in hidden Markov(-switching) models. The Kim filter is extended for novelty, enabling parameter and latent variable estimation within a frequentist counterpart to the NDLC-SEM framework. Similar to \cite{Kelava2022}'s, our approach accommodates unobserved heterogeneity in the hidden Markov(-switching) models. In the present model, the nonlinearity takes the form of interactions between lagged intra-individual states and inter-individual traits, which enter in two places: the person-specific autoregressive coefficients of the within-level dynamics (through $\bm{B}_{4s}$) and the regime-transition probabilities (through $\bm{\gamma}_{4}$). This differs from the nonlinear state-space dynamics in \citet{Chow2013}, which arise from nonlinear state equations.

Figure~\ref{fig:model_schematic} provides a schematic overview of the proposed RSSS model. The within-person level (top) shows the dynamic latent variables $\bm{\eta}_{1it}$ and their regime-dependent VAR(1) dynamics, together with the discrete latent regime variable $S_{it}$. The between-person level (bottom) shows the time-invariant latent variables $\bm{\eta}_{2i}$, which influence both the within-person dynamics and the transition probabilities between regimes. The full model specification is given in Section~\ref{sec:models}.

\begin{figure}[H]
\centering
\begin{tikzpicture}[
  latent/.style={circle, draw, minimum size=1.35cm, inner sep=1pt, font=\small},
  observed/.style={rectangle, draw, minimum size=0.85cm, inner sep=3pt, font=\small},
  regime/.style={circle, draw, dashed, minimum size=1.2cm, inner sep=1pt, font=\small},
  >=stealth
]
\node[latent]   (eta1l) at (3.0, 5.0) {$\bm{\eta}_{1i,t{-}1}$};
\node[observed] (y11l)  at (1.2, 7.3) {\scriptsize $y_{11,i,t{-}1}$};
\node[observed] (y12l)  at (3.0, 7.3) {\scriptsize $y_{12,i,t{-}1}$};
\node[observed] (y13l)  at (4.8, 7.3) {\scriptsize $y_{13,i,t{-}1}$};
\draw[->] (eta1l) -- (y11l);
\draw[->] (eta1l) -- (y12l);
\draw[->] (eta1l) -- (y13l);
\node[latent]   (eta1r) at (9.0, 5.0) {$\bm{\eta}_{1it}$};
\node[observed] (y11r)  at (7.2, 7.3) {\scriptsize $y_{11,it}$};
\node[observed] (y12r)  at (9.0, 7.3) {\scriptsize $y_{12,it}$};
\node[observed] (y13r)  at (10.8,7.3) {\scriptsize $y_{13,it}$};
\draw[->] (eta1r) -- (y11r);
\draw[->] (eta1r) -- (y12r);
\draw[->] (eta1r) -- (y13r);
\draw[->] (eta1l) -- node[above, font=\scriptsize] {$\bm{B}_{3i,S_{i,t-1}}$} (eta1r);
\node[regime] (Sl) at (5.0, 3.5) {$S_{i,t{-}1}$};
\node[regime] (Sr) at (9.0, 3.5) {$S_{it}$};
\draw[->] (Sl) -- (Sr);
\node[font=\scriptsize, right=3pt of Sr] {$\sigma(\cdot)$};
\draw[->] (eta1l) to[out=-30, in=160] node[pos=0.45, above, font=\scriptsize] {$\bm{\gamma}_3$} (Sr);
\draw[->, dashed] (eta1l) .. controls (3.5, 2.1) and (7.5, 2.1) .. node[pos=0.5, above=1pt, font=\scriptsize] {$\bm{\gamma}_4$} (Sr);
\draw[dashed] (-0.3, 2.0) -- (11.5, 2.0);
\node[font=\small, anchor=west] at (11.2, 2.4) {within};
\node[font=\small, anchor=west] at (11.2, 1.6) {between};
\node[latent]   (eta2) at (5.0,  0.3) {$\bm{\eta}_{2i}$};
\node[observed] (y21)  at (1.5,  1.1) {\scriptsize $y_{21i}$};
\node[observed] (y22)  at (1.5,  0.3) {\scriptsize $y_{22i}$};
\node[observed] (y23)  at (1.5, -0.5) {\scriptsize $y_{23i}$};
\draw[->] (eta2) -- (y21);
\draw[->] (eta2) -- (y22);
\draw[->] (eta2) -- (y23);
\draw[->] (eta2) -- (eta1l);
\draw[->] (eta2) to[bend right=15] (eta1r);
\draw[->] (eta2) to[out=30, in=250] node[pos=0.55, right, font=\scriptsize] {$\bm{\gamma}_2$} (Sr);
\draw[->, dashed] (eta2) to[out=60, in=220] (Sr);
\end{tikzpicture}%
\caption{Schematic of the regime-switching state-space (RSSS) model. Solid circles represent continuous latent variables; rectangles represent observed variables; dashed circles represent the discrete latent regime $S_{it} \in \{1,2\}$. The dashed horizontal line separates the within-person level (top) from the between-person level (bottom). The solid arrow from $\bm{\eta}_{1i,t-1}$ to $\bm{\eta}_{1it}$ (labelled $\bm{B}_{3i,S_{i,t-1}}$) represents the regime-dependent VAR(1) dynamics. Solid arrows from $\bm{\eta}_{2i}$ to the within-person level represent effects on intercepts ($\bm{B}_{2s}$) and autoregressive coefficients ($\bm{B}_{4s}$). The two solid arrows pointing to $S_{it}$ represent main effects on transition probabilities ($\bm{\gamma}_{2}$ from $\bm{\eta}_{2i}$; $\bm{\gamma}_{3}$ from $\bm{\eta}_{1i,t-1}$); the two dashed arrows pointing to $S_{it}$ (from $\bm{\eta}_{1i,t-1}$ and from $\bm{\eta}_{2i}$) jointly represent the $\bm{\eta}_{1i,t-1} \times \bm{\eta}_{2i}$ interaction effect ($\bm{\gamma}_{4}$)}
\label{fig:model_schematic}
\end{figure}

The remainder of this article is structured as follows. In the next section we introduce the RSSS modeling framework, followed by a detailed description of the model estimation and the implementation of the extended filter. In an empirical example, we illustrate the approach and its ability to forecast regime membership. Then, we provide further evidence of its performance using a simulation study. We end with a discussion of the strengths and weaknesses of the approach.

\section{Motivating Example}\label{sec:motiv}

The empirical example that motivates the present work is the SAM study (a German acronym for university dropout in mathematics) \citep{Kelava2022}, a longitudinal investigation of university student dropout in mathematics at a German university. It is well documented that approximately 40\% of students drop out in the early phase of German mathematics programs \citep{Heublein2014}. Understanding the process leading to dropout requires modeling both relatively stable between-individual characteristics (e.g., cognitive skills) and rapidly fluctuating within-individual affective states (e.g., stress, fear of failure), as well as their interplay over time.

Following the Rubicon model of action phases \citep{Achtziger2018}, we conceptualize the transition from \textit{no dropout intention} to \textit{dropout intention} as a discrete latent regime switch. Once a student crosses this psychological threshold, the decision is treated as essentially irreversible in the short term. This motivates a regime-switching state-space (RSSS) model in which the latent regime $S_{it} \in \{1, 2\}$ evolves as a first-order Markov process, with $S_{it} = 1$ denoting no dropout intention and $S_{it} = 2$ denoting dropout intention.

From a substantive perspective, several modeling features of the empirical application are of particular interest. First, affective states such as \textit{cost} (perceived cost of studying) and \textit{afraid to fail} (fear of failure) fluctuate within individuals over time and are expected to influence the likelihood of forming a dropout intention. These are represented by the intra-individual latent factor vector $\bm{\eta}_{1it}$ (comprising all seven affective scales), whose lagged values $\bm{\eta}_{1i,t-1}$ enter the transition probabilities. Second, between-individual differences in cognitive skills $\bm{\eta}_{2i}$ are expected to moderate the relationship between affective states and dropout intention: a student with higher cognitive resources may be better protected against the effect of momentary stress or fear of failure on regime switching. This motivates the transition probability model
\begin{equation*}
    \Pr[S_{it}=1 \mid S_{i,t-1}=s',\, \bm{\eta}_{1i,t-1},\, \bm{\eta}_{2i}]
    = \sigma\!\left(
        \gamma_{s'1}
        + \gamma_{s'2}\,\bm{\eta}_{2i}
        + \bm{\gamma}_{s'3}\,\bm{\eta}_{1i,t-1}
        + \bm{\gamma}_{s'4}\,\bm{\eta}_{1i,t-1}\bm{\eta}_{2i}
    \right),
\end{equation*}
where $\sigma(x)=[1+\exp(-x)]^{-1}$ is the sigmoid function; $\gamma_{s'2}$ captures the main effect of cognitive skills on the probability of remaining in the no-dropout-intention regime; $\bm{\gamma}_{s'3}$ captures the within-individual effects of the seven affective states; and $\bm{\gamma}_{s'4}$ captures how cognitive skills moderate those effects (see Equation \ref{MS} for the full formal specification). Together, these features call for a frequentist estimation procedure that can handle latent within- and between-individual predictors of regime transitions: the main contribution of the present work.

The dataset contains three types of information: a questionnaire on the inter-individual differences, a short online survey on intra-individual differences collected over 50 time points, and attendance information for the accompanying tutorial sessions. Our sample contains 119 mathematics students at a German university. Seven intra-individual latent factors (\textit{content not important, cost, afraid to fail, not understanding, stress, positive affect (reversed), negative affect}) are estimated from 17 observed variables, and 1 inter-individual latent factor (\textit{cognitive skills}) is estimated from 2 items.\footnote{3 items were used in the original study \citep{Kelava2022}; however, one of the items had a weak correlation with others and was not useful; thus it was omitted from the analysis.} Table \ref{tab:latent_factors} briefly summarizes the meaning of the intra- and inter-individual factors.

In the dataset, 14 individuals (11.8\%) dropped out by the middle time point ($t=25$) and 45 individuals (37.8\%) dropped out by the final time point ($t=50$). The intra-individual variables ($\bm{y}_{1}$) had a large proportion of missing values. Among 119 $\times$ 50 observations, 55.8\% of the entries had at least one variable missing before the dropout and 88.2\% after the dropout.

\begin{table}
\centering
\caption{Latent factors considered in the empirical study and their corresponding observed variables used for measurement}
\begin{tabular}{@{}ll@{}}
\toprule
\textbf{Factor} & \textbf{Item Description} \\ \midrule
\textbf{Content not important (3 items)} & Low attainment/utility value. \\
\textbf{Cost (2 items)} & Perceived cost of studying. \\
\textbf{Afraid to fail (2 items)} & Fear of failure and thoughts of dropping out. \\
\textbf{Not understanding (2 items)} & Lack of understanding of course content and tasks. \\
\textbf{Stress (2 items)} & Stress and being overwhelmed by the course demands. \\
\textbf{Positive affect (3 items)} & PANAS positive scale (reversed). \\
\textbf{Negative affect (3 items)} & PANAS negative scale. \\
\textbf{Cognitive skills (2 items)} & Mathematics grade from high school (Abitur). \\
& Sum score of TIMSS mathematics items. \\ \bottomrule
\end{tabular}
\label{tab:latent_factors}
\end{table}

\section{Regime-Switching State-Space (RSSS) models}\label{sec:models}
In this section, we provide a brief description of our models that govern the underlying data generating process. The measurement model establishes the relationship between the latent variables and their observed proxies. The structural model depicts the temporal progression of continuous intra-individual latent variables through hidden Markov 1st order vector autoregressive (VAR(1)) models, which depend on both inter- and intra-individual variables. The Markov-switching model illustrates the regime-switching mechanisms through transition probabilities. The transition probabilities are described as a function of intra- and inter-individual latent variables as well as their interactions. Lastly, actual regime-switch occurs as a random event based on the transition probabilities.

\subsection{Measurement model}
The measurement model is described by the following linear equations. Dynamic observed variables, which are expressed as $(\bm{y}_{1it} | S_{it}=s)$, are conditional on individual $i$, time $t$, and a discrete latent state $S_{it}$. Dynamic latent variables, which are expressed as $(\bm{\eta}_{1it} | S_{it}=s)$ are also conditional on individual $i$, time $t$, and a discrete latent state $S_{it}$,
\begin{equation} \label{MM1}
    (\bm{y}_{1it} | S_{it}=s) = \bm{\Lambda}_{1s}(\bm{\eta}_{1it} | S_{it}=s) + \bm{\epsilon}_{1its},
\end{equation}
with a conditional vector of uncorrelated residuals $\bm{\epsilon}_{1its} \sim \mathcal{N}(\bm{0}, \bm{R}_{1s})$. Let $O_1$ be the number of time-dependent observed variables and $U_1$ be the number of time-dependent latent variables.  $\bm{\Lambda}_{1s}$ is a factor loading matrix which is dependent on dynamic latent state $S_{it}$, with a dimension $O_1 \times U_1$. $\bm{R}_{1s}$ is an $O_1 \times O_1$ residual covariance matrix which is dependent on a dynamic latent state $S_{it}$.

Time-invariant observed variables ($\bm{y}_{2i}$) and time-invariant latent variables ($\bm{\eta}_{2i}$) are dependent on individual $i$,
\begin{equation} \label{MM2}
    \bm{y}_{2i} =
    \bm{\Lambda}_{2} \bm{\eta}_{2i} +           \bm{\epsilon}_{2i},
\end{equation}
where $\bm{\eta}_{2i} \sim \mathcal{N}(\bm{0}, \bm{P}_{2}), \bm{\epsilon}_{2i} \sim \mathcal{N}(\bm{0}, \bm{R}_2)$ are vectors of residuals. Let $O_2$ be the number of time-invariant observed variables and $U_2$ be the number of time-invariant latent variables. $\bm{\Lambda}_{2}$ is a $O_2 \times U_2$ factor loading matrix. $\bm{P}_{2}$ is a $U_2 \times U_2$ factor covariance matrix and $\bm{R}_2$ is an $O_2 \times O_2$ residual covariance matrix.

\subsection{Within-level structural model}
The within-level structural model is described by the conditional linear equation given a temporal discrete latent state $S_{it}=s$,
\begin{equation} \label{WSM}
    (\bm{\eta}_{1it} | S_{it}=s) = \bm{b}_{1is} + \bm{B}_{3is} \bm{\eta}_{1i,t-1} + \bm{\zeta}_{1its},
\end{equation}
where $\bm{\zeta}_{1its} \sim \mathcal{N}(\bm{0}, \bm{Q}_{1s})$. For person $i$ at time $t$, $(\bm{\eta}_{1it} | S_{it}=s)$ represents a $U_1 \times 1$ vector of conditional intra-individual latent scores.
$\bm{b}_{1is}$ is a $U_1 \times 1$ conditional random intercept vector, $\bm{B}_{3is}$ is a $U_1 \times U_1$ matrix of conditional intra-individual VAR(1) effects, and $\bm{Q}_{1s}$ is a $U_1 \times U_1$ residual covariance matrix which is dependent on a dynamic latent state $S_{it}$.

\subsection{Between-level structural model}
For the random intercept vector $\bm{b}_{1is}$ and the autoregressive coefficient matrix $\bm{B}_{3is}$, the between-level structural model is specified by two models:
\begin{align} \label{BSM}
    \bm{b}_{1is} &= \bm{b}_{1s} + \bm{B}_{2s} \bm{\eta}_{2i} + \bm{\zeta}_{2i}, \\
    \bm{B}_{3is} &= \bm{B}_{3s} + \bm{B}_{4s} \bm{\eta}_{2i}, \label{BSM2}
\end{align}
where $\bm{b}_{1s}$ is a $U_1 \times 1$ conditional intercept vector, $\bm{B}_{2s}$ is a $U_1 \times U_2$ matrix of conditional inter-individual effects, $\bm{\zeta}_{2i}$ is a $U_1 \times 1$ vector of random intercepts with $\bm{\zeta}_{2i} \sim \mathcal{N}(\bm{0}, \bm{Q}_{2})$. $\bm{Q}_2$ is a $U_1 \times U_1$ residual covariance matrix. $\bm{B}_{3s}$ is a $U_1 \times U_1$ matrix of conditional intra-individual VAR(1) effects, and $\bm{B}_{4s}$ is a $U_1 \times U_1 \ \times \ U_2$ matrix of conditional interaction effects.\footnote{$\bm{B}_{4s}$ is a $U_1 \times U_1 \times U_2$ array; the product $\bm{B}_{4s}\bm{\eta}_{2i}$ denotes the sum $\sum_{k=1}^{U_2} \bm{B}_{4s}^{(k)}\,\eta_{2i,k}$, where $\bm{B}_{4s}^{(k)}$ is the $k$-th $U_1 \times U_1$ slice of the array and $\eta_{2i,k}$ is the $k$-th element of $\bm{\eta}_{2i}$, so that the result is a $U_1 \times U_1$ matrix, conformable with $\bm{B}_{3s}$ in Equation~\eqref{BSM2}. $\bm{B}_{4s}$ thereby enters both the between-level structural model (multiplying $\bm{\eta}_{2i}$) and, through $\bm{B}_{3is}$, the within-level structural model (multiplying $\bm{\eta}_{1i,t-1}$).} Individual differences in autoregressive dynamics are captured entirely through the structured term $\bm{B}_{4s}\bm{\eta}_{2i}$, following the NDLC-SEM specification of \citet{Kelava2022}, in which the autoregressive coefficient is modeled as a deterministic function of the between-person factor while the random-effect term enters only the intercept equation; no additional unstructured random-effect term is included in Equation~\eqref{BSM2}.

The between-level specification of $\bm{B}_{3is}$ in Equation~\ref{BSM2} is related to, but distinct from, formulations in DSEMs and mlVAR. In those frameworks, the person-specific autoregressive matrix is typically parameterized as $\bm{B}_{3is} = \bm{B}_{3s} + \bm{u}_i$, where $\bm{u}_i$ is a vector of unconstrained person-specific random effects, one for each unique element of $\bm{B}_{3is}$. In the present model, the analogous individual deviation is instead expressed as $\bm{B}_{4s}\bm{\eta}_{2i}$, in which $\bm{\eta}_{2i}$ is a lower-dimensional vector of between-person latent factors that simultaneously accounts for covariation in the time-invariant observed variables $\bm{y}_{2i}$ (Equation~\ref{MM2}) and for individual differences in autoregressive dynamics. This formulation is more parsimonious when between-person differences in dynamics can be attributed to a small number of latent traits. As a special case, if $U_2$ equals the number of unique parameters in $\bm{B}_{3is}$ (one latent factor per autoregressive parameter), then the proposed model reduces to the DSEM/mlVAR parameterization: $\bm{B}_{4s}\bm{\eta}_{2i}$ spans the same space as the unconstrained random-effects vector $\bm{u}_i$, and the two formulations are equivalent.

A natural question is whether the latent predictor $\bm{\eta}_{2i}$ could be replaced by its observed proxy $\bm{y}_{2i}$ directly. A latent-variable representation is theoretically preferable in at least three distinct situations, each addressing a different problem. First (dimensionality), when the number of between-person indicators $O_2$ is large, entering each $y_{2j}$ separately as a predictor introduces $O_2$ regression coefficients for a single construct, creating multicollinearity among correlated indicators; the factor analysis underlying $\hat{\bm{\eta}}_{2i}$ achieves dimensionality reduction, replacing $O_2$ collinear predictors with a single (or low-dimensional) latent score. Second (bias in the effect estimate), when the measurement error variance substantially exceeds the factor variance ($\bm{R}_2 \gg \bm{P}_2$), the observed scores are dominated by noise, so using them directly attenuates the regression coefficient toward zero by a factor equal to the reliability $\lambda = P_2/(P_2 + R_2)$, a bias that becomes severe as $\bm{R}_2$ grows relative to $\bm{P}_2$. Importantly, the corresponding de-attenuation does not require $\bm{R}_2$ itself to be estimated precisely, since the Bartlett factor score is conditionally unbiased for $\bm{\eta}_{2i}$ under any weighting matrix used in its computation. Third (bias in the variance decomposition), when $\bm{\eta}_{2i}$ carries its own variance component $\bm{P}_2$, treating it as latent keeps genuine between-person variance separate from measurement error, whereas collapsing it onto the observed proxy conflates the two \citep[cf.][p.~742, fn.~1]{Chow2013}. Conversely, if $\bm{P}_2$ is negligible or measurement error in $\bm{y}_{2i}$ is small, observed predictors offer a simpler and equally defensible alternative.

\subsection{Markov-switching model}
The transition probabilities are characterized using a nonlinear function of the latent variables of the between- and within-level structural models:
\begin{align} \label{MS}
    Pr[S_{it}=1 | S_{i,t-1}=s'] &= \sigma(\gamma_{s'1} +  \bm{\gamma}_{s'2} \bm{\eta}_{2i} + \bm{\gamma}_{s'3} \bm{\eta}_{1i,t-1} + \bm{\gamma}_{s'4} \bm{\eta}_{1i,t-1} \bm{\eta}_{2i}), \\
    Pr[S_{it}=2 | S_{i,t-1}=s'] &= 1 - Pr[S_{it}=1 | S_{i,t-1}=s'],
\end{align}
where $s,s' \in \{1,2\}$ indicates the regime membership (e.g., $s=1$: no dropout intention; $s=2$: dropout intention). $\sigma$ represents the sigmoid function i.e., $\sigma(x) = \left[1 + \exp(-x) \right]^{-1}$. $\gamma_{s'1}$ is an intercept, $\bm{\gamma}_{s'2}$ is a $1 \times U_2$ vector corresponding to the effect of an inter-individual factor, $\bm{\gamma}_{s'3}$ is a $1 \times U_1$ vector of the effects of intra-individual factors, and $\bm{\gamma}_{s'4}$ is a $1 \times U_1$ vector of the interaction effects. The underlying regime-switching mechanism is described as a random event from the probability distribution  $Pr[S_{it}=s | S_{i,t-1}=s']$. From a psychometric perspective, the Markov-switching component can be viewed as an intensive-longitudinal analogue of latent transition modeling \citep{CollinsWugalter1992,CollinsLanza2010}: the discrete regime variable $S_{it}$ plays the role of a latent status at each time point, and Equation~\ref{MS} specifies the first-order transition probabilities between those statuses. Unlike conventional latent transition models, however, the present framework embeds these transitions in a regime-switching state-space model and allows the transition probabilities to depend on time-varying within-person latent states, time-invariant between-person latent traits, and their interaction. Equation~\ref{MS} also nests simpler transition specifications as special cases: setting $\bm{\gamma}_{s'2} = \bm{\gamma}_{s'3} = \bm{\gamma}_{s'4} = \bm{0}$ yields transition probabilities that are constant across persons, as in the nonlinear RSSS formulation of \citet{Chow2013}, while restricting only $\bm{\gamma}_{s'3}$ and $\bm{\gamma}_{s'4}$ to zero yields a model in which transitions depend on between-person differences alone. Applied users can therefore move between these specifications within the same estimation routine and assess empirically whether the additional latent predictors are supported by their data.

\section{Parameter Estimation and Forecasting}\label{sec:est}
The proposed parameter estimation procedure is summarized in Algorithm 1. The algorithm computes the marginal maximum likelihood of the structural parameters $\bm{\theta}$, in which the person-specific random intercept residuals $\bm{\zeta}_{2i}$ are marginalized out within the filter under their model-implied distribution $\mathcal{N}(\bm{0}, \bm{Q}_2)$ (Section~\ref{sec:integrated}); the variance component $\bm{Q}_2$ is an element of $\bm{\theta}$ and is estimated jointly with all other parameters. In step~1, initial state moments and parameters are initialized; the initial distribution of the augmented state includes the $\bm{\zeta}_{2i}$ block at its $\mathcal{N}(\bm{0}, \bm{Q}_2)$ distribution (Section~\ref{sec:integrated}). In step~2, factor scores of inter-individual latent variables $\hat{\bm{\eta}}_{2i}$ are calculated via confirmatory factor analysis.
In steps~3--5, the extended Kim filter, which is composed of the Kalman filter, the extended Hamilton filter, and the collapsing process, is used to filter RSSS models on the original observations $\bm{y}_{1it}$, operating on the augmented state that carries the $\bm{\zeta}_{2i}$ block.
In step~6, the prediction errors, which follow a multivariate normal distribution under the distributional assumptions, serve as a basis for parameter updates of $\hat{\bm{\theta}}$ using an approximated likelihood function.\footnote{The likelihood is approximated in the collapsing procedure \citep{Kim&Nelson1999} and can be written using the predictive error decomposition function \citep{Schweppe1965}.} Given the set of parameter values at time $t$, the extended Kim filter forecasts the dynamic latent states and the underlying regime memberships for time $t+1$. The parameter estimates will become available at the end of Algorithm 1; the associated standard errors can be obtained by taking the square root of the diagonal elements of the negative numerical Hessian matrix or its approximation at the estimated point.

\begin{algorithm}
\caption{Marginal maximum likelihood via the extended Kim filter} \label{algo1}
\begin{algorithmic}
    \STATE \text{1: } Initialization ($\bm{\zeta}_{2i}$ block at $\mathcal{N}(\bm{0}, \bm{Q}_2)$)
    \WHILE{not converged}
        \STATE \text{2: } Confirmatory factor analysis
        \FOR{$t=1,\ldots,T$}
            \STATE \text{3: } Kalman filter (augmented state)
            \STATE \text{4: } Extended Hamilton filter
            \STATE \text{5: } Collapsing process
        \ENDFOR
        \STATE \text{6: } Parameter update ($\hat{\bm{\theta}}$, including $\hat{\bm{Q}}_2$)
    \ENDWHILE
\end{algorithmic}
\end{algorithm}

\subsection{Initialization (step 1)}\label{sec:est:init}
While the frequentist approach avoids the need to specify priors, it still requires starting values to initialize the Kim filter.
Let $\mathcal{D}_{1:t} =\{\bm{y}_{1i1:t}, \bm{y}_{2i} \}_{i=1}^{N}$ be the collection of available data at time $t$, $\bm{\eta}_{1it|t-1}^{s,s'} = \mathbb{E}(\bm{\eta}_{1it} | S_{it}=s, S_{i,t-1}=s', \mathcal{D}_{1:t-1})$ be the conditional mean vector of intra-individual latent variables, $\bm{P}_{1it|t-1}^{s,s'} = \text{Cov}(\bm{\eta}_{1it} | S_{it}=s, S_{i,t-1}=s', \mathcal{D}_{1:t-1})$ be the associated covariance matrix, the dynamic latent factor scores at $t=0$ are assumed to be normally distributed, i.e., $\bm{\eta}_{1i0} \sim \mathcal{N}(\bm{\eta}_{1i0|0}, \bm{P}_{1i0|0})$. Typically, $\bm{\eta}_{1i0}$ is assumed to have a \textit{diffuse} density, implying that $\bm{\eta}_{1i0|0}$ is fixed to a vector of constants (e.g., a vector of zeros) and the diagonal elements of the covariance matrix $\bm{P}_{1i0|0}$ are set to arbitrarily large constants. Specific initialization choices are described in Section~\ref{sec:emp}. More principled alternatives include model-implied initial moments derived from the stationary distribution of the state process \citep{Harvey1990, duToit2007} and, for the initial regime probabilities, from the ergodic distribution of the Markov transition matrix \citep{Yang2010}. In a regime-switching model, however, these model-implied moments admit more than one construction, and the existing literature does not settle the choice. \citet{Chow2013} initialize the state with a single regime-independent distribution (a zero mean vector and an identity covariance matrix in their simulation study) and reserve the model-implied route for the initial regime probabilities, which they obtain from the transition probability parameters \citep[cf.][]{Yang2010}; model-implied stationary state moments are raised in their discussion as a recommendation for linear stationary models rather than implemented. A regime-averaged construction is also possible in principle. Because no data are available at $t=0$, the mixing weights cannot be filtered probabilities and would have to be taken from the ergodic distribution $\bm{\pi}$ of the transition matrix, so that $\bm{\eta}_{1i0}$ is initialized at the moments of the mixture $\sum_{s} \pi_{s}\,\mathcal{N}(\bm{\eta}^{(s)}_{1i0|0}, \bm{P}^{(s)}_{1i0|0})$. We instead apply the stationary moments \emph{conditionally on each regime}, using that regime's own stationary distribution. For the diagonal AR(1) specification (Assumption~4 in Section~\ref{sec:emp:impl}) both regime-conditional stationary moments take a simple closed form: the mean $\bm{\eta}^{(s)}_{1i0|0} = (\bm{I}-\bm{B}_{3is})^{-1}\bm{b}_{1is}$, which reduces element-wise to $(\bm{b}_{1is})_{j}/(1-(\bm{B}_{3is})_{jj})$, and the covariance $\bm{P}^{(s)}_{1i0|0} = \operatorname{diag}\!\big((\bm{Q}_{1})_{jj}/(1-(\bm{B}_{3is})_{jj}^{2})\big)$. The covariance depends only on the within-regime autoregressive dynamics and process noise, whereas the mean additionally involves the regime-specific intercept $\bm{b}_{1is}$; both are well defined under the stationarity condition $|(\bm{B}_{3is})_{jj}|<1$ of Assumption~4. The choice between the regime-conditional and the regime-averaged construction, and the applicability of the ergodic initialization of the regime probabilities, both depend on the transition structure of the specific application; when one regime is close to absorbing, $\bm{\pi}$ concentrates on that regime and the regime-averaged construction, while still well defined, ceases to be substantively appropriate. These considerations are taken up in Section~\ref{sec:emp:impl}. Under the marginal formulation of Section~\ref{sec:integrated}, the state vector additionally carries the time-invariant block $\bm{\zeta}_{2i}$, and the initialization extends to the joint stationary distribution of $(\bm{\eta}_{1i0}, \bm{\zeta}_{2i})$: the $\bm{\zeta}_{2i}$ block is initialized at its model-implied distribution $\mathcal{N}(\bm{0}, \bm{Q}_2)$, and the cross-covariance between the two blocks is $(\bm{I}-\bm{B}_{3i1})^{-1}\bm{Q}_2$, as implied by the intercept equation (see Section~\ref{sec:integrated}). Regarding parameter starting values, the optimization is performed with multiple random initializations, and the solution with the highest log-likelihood is retained, which mitigates sensitivity to any single starting value.

\subsection{The Confirmatory factor analysis (step 2)}
The inter-individual latent variables $\bm{\eta}_{2i}$ are estimated from their observed proxies $\bm{y}_{2i}$ using the factor score regression \citep[cf.,][] {Skrondal2001}. The calculation follows the standard formula:

\begin{equation} \label{F2}
    \hat{\bm{\eta}}_{2i} = \bm{F}_{2} \bm{y}_{2i}
\end{equation}
where $\bm{F}_{2}$ is the matrix of factor score weights. This matrix is computed based on the current estimates of the measurement model parameters: the factor loading matrix $\bm{\Lambda}_2$ and the residual covariance matrix $\bm{R}_2$. In the case of Bartlett factor scores \citep{Bartlett1937}, the formula for $\bm{F}_{2}$ is:

\begin{equation}
    \bm{F}_{2} = \left( \bm{\Lambda}_2^T \bm{R}_2^{-1} \bm{\Lambda}_2 \right)^{-1} \bm{\Lambda}_2^T \bm{R}_2^{-1}.
\end{equation}

The Bartlett factor scores provide a conditionally unbiased estimate of the factor scores given the observed data \citep{Bartlett1937}; see also \citet{Chow2010} for the well-known correspondence between regression/Bartlett estimators of factor scores and the Kalman filter. Bartlett factor scores are conditionally unbiased for the latent variables, whereas regression factor scores are shrunk toward the population mean \citep{Skrondal2001}. For a single linear regression among factors, \citet{Skrondal2001} show that this shrinkage is precisely what makes regression scores unbiased as exogenous predictors; that recipe does not carry over to the present setting, where $\hat{\bm{\eta}}_{2i}$ enters several equations simultaneously (the intercept and the autoregressive modulation of the within-level dynamics, and the nonlinear transition model) and where the regression score weights would require the factor variance $P_2$, which is itself estimated within the iteration. We therefore use Bartlett scores as plug-in estimates of the between-person latent variables in the subsequent filtering procedure: their classical error structure (mean zero given the true factor, for any value of $\bm{R}_2$ used in the weights) keeps the consequences of the plug-in analyzable, and the resulting attenuation is quantified in Section~\ref{sec:sim}. 

\subsection{The Kalman filter (step 3)}

The original Kalman filter \citep{Kalman1960} was proposed as a real-time estimation procedure to estimate the continous latent variables in a state-space model as well as to predict their future values. The Kalman filter consists of three steps, namely, \textit{prediction},  \textit{measurement}, and \textit{update}.

In step 3.1, a prediction of the within-level latent variables $\bm{\eta}_{1it}$ is made based on all available information up to the previous time point, denoted $\bm{\eta}_{1it|t-1}$. In step 3.2, the \textit{one-step-ahead prediction error} is calculated as the difference between the observed variables $\bm{y}_{1it}$ and their prediction. In step 3.3, the prediction $\bm{\eta}_{1it|t}$ is updated based on the likelihood. These three steps are repeated at each measurement occasion. The exact reformulation of the state-space model is specified in the appendix.

Using the Kalman filter, we derive dynamic latent factor scores in real-time as a new observation arrives. Let
$\bm{v}_{it}^{s,s'}$ be the conditional one-step-ahead prediction error vector of intra-individual observed variables, and $\bm{F}_{1it}^{s,s'}$ the associated covariance matrix. Here, the set ($s$, $s'$) represent indices of the current and previous regimes. The Kalman filter procedure consists of computing the following quantities in order:
\begin{align} \label{Kalman1}
    \bm{\eta}_{1it|t-1}^{s,s'} &= \bm{b}_{1is} + \bm{B}_{3is} \bm{\eta}_{1i,t-1|t-1}^{s'} \\
 \label{Kalman2}
    \bm{P}_{1it|t-1}^{s,s'} &= \bm{B}_{3is} \bm{P}_{1i,t-1|t-1}^{s'} \bm{B}_{3is}^{T} + \bm{Q}_{1s} \\
 \label{Kalman3}
    \bm{v}_{it}^{s,s'} &= \bm{y}_{1it} - \bm{\Lambda}_{1s} \bm{\eta}_{1it|t-1}^{s,s'}\\
 \label{Kalman4}
    \bm{F}_{1it}^{s,s'} &= \bm{\Lambda}_{1s} \bm{P}_{1it|t-1}^{s,s'} \bm{\Lambda}_{1s}^T + \bm{R}_{1s}\\
 \label{Kalman5}
    \bm{\eta}_{1it|t}^{s,s'} &= \bm{\eta}_{1it|t-1}^{s,s'} + \bm{K}_{it}^{s,s'} \bm{v}_{it}^{s,s'}\\
 \label{Kalman6}
    \bm{P}_{1it|t}^{s,s'} &= \bm{P}_{1it|t-1}^{s,s'} - \bm{K}_{it}^{s,s'} \bm{\Lambda}_{1s} \bm{P}_{1it|t-1}^{s,s'}
\end{align}
where $\bm{b}_{1is} = \bm{b}_{1s} + \bm{B}_{2s} \hat{\bm{\eta}}_{2i}$ is the person- and regime-specific intercept absorbing the between-level factor scores (step~2); the random intercept residual $\bm{\zeta}_{2i}$ is not part of this intercept but is carried as a block of the augmented state vector and marginalized within the filter (Section~\ref{sec:integrated}). For notational economy, Equations~(\ref{Kalman1}) to (\ref{Kalman7}) are written for the within-level block $\bm{\eta}_{1it}$; in the implementation they operate on the augmented state $\bm{x}_{it} = (\bm{\eta}_{1it}^{\top}, \bm{\zeta}_{2i}^{\top})^{\top}$ with the transition and measurement matrices given in Section~\ref{sec:integrated}, and $\bm{K}_{it}^{s,s'} = \bm{P}_{1it|t-1}^{s,s'} \bm{\Lambda}_{1s}^{T} [\bm{F}_{1it}^{s,s'}]^{-1}$ is called the Kalman gain function. Equation (\ref{Kalman6}) is often substituted by what's called the \textit{Joseph form} \citep{Bucy2005} because subtracting two covariance matrices may result in a numerical error causing one of the diagonal elements to be slightly less than zero. This poses a serious issue as it violates the positive (semi-)definiteness of the covariance matrix. To circumvent this issue, the \textit{Joseph form} update can be employed. By rearranging Equation (\ref{Kalman6}), we obtain
\begin{equation}  \label{Kalman7}
    \bm{P}_{1it|t}^{s,s'} = (\bm{I} - \bm{K}_{it}^{s,s'} \bm{\Lambda}_{1s}) \bm{P}_{1it|t-1}^{s,s'} (\bm{I} - \bm{K}_{it}^{s,s'} \bm{\Lambda}_{1s})^{T} + \bm{K}_{it}^{s,s'} \bm{R}_{1s} (\bm{K}_{it}^{s,s'})^T,
\end{equation}
where $\bm{I}$ represents an identity matrix (of size $U_1$). The Kalman filter summarized in Equations (\ref{Kalman1}) to (\ref{Kalman7}) operates recursively for $t=1,\ldots,T$ until $\bm{\eta}_{1it|t}^{s,s'}$ and $\bm{P}_{1it|t}^{s,s'}$ have been computed for all time points.

\subsection{The extended Hamilton filter (step 4)}

The Hamilton filter \citep{hamilton1989} is an algorithm to make inference in the Markov-switching model. Unlike the Kalman filter, the algorithm is specialized for the inference of discrete latent states and works as a nonlinear filter. The filter also utilizes the maximum likelihood estimation which serves as a basis for parameter optimization. Similar to the Kalman filter, the Hamilton filter consists of a \textit{prediction} and an \textit{update} step.

In step 4.1, the model probabilities are predicted using the transition probabilities. In step 4.2, the prediction is updated based on the new observations. The model probabilities are then reweighted by the likelihood.

Our version is an extension to the original Hamilton filter \citep{hamilton1989}, accounting for within- and between-individual differences in the regime-switching trajectory. We call it the extended Hamilton filter and the recursive process can be described as
\begin{align} \label{HF1}
    &Pr[S_{it}=s, S_{i,t-1}=s'| \mathcal{D}_{1:t-1}] = Pr[S_{it}=s| S_{i,t-1}=s'] Pr[S_{i,t-1}=s'| \mathcal{D}_{1:t-1}], \\
    &f(\bm{y}_{1it}| \mathcal{D}_{1:t-1}) = \sum_{s,s'} f(\bm{y}_{1it}| S_{it}=s, S_{i,t-1}=s', \mathcal{D}_{1:t-1}) Pr[S_{it}=s, S_{i,t-1}=s'| \mathcal{D}_{1:t-1}] \label{HF2}\\
    & Pr[S_{it}=s, S_{i,t-1}=s'| \mathcal{D}_{1:t}] =
    \frac{f(\bm{y}_{1it}| S_{it}=s, S_{i,t-1}=s', \mathcal{D}_{1:t-1}) Pr[S_{it}=s, S_{i,t-1}=s'| \mathcal{D}_{1:t-1}]}{f(\bm{y}_{1it}| \mathcal{D}_{1:t-1})} \label{HF3}\\
    & Pr[S_{it}=s| \mathcal{D}_{1:t}] = \sum_{s'} Pr[S_{it}=s, S_{i,t-1}=s' | \mathcal{D}_{1:t}], \label{HF4}
\end{align}
where the conditional multivariate normal likelihood function $f(\bm{y}_{1it}| S_{it}=s, S_{i,t-1}=s', \mathcal{D}_{1:t-1})$ is computed as
\begin{equation} \label{lik}
    f(\bm{y}_{1it}| S_{it}=s, S_{i,t-1}=s', \mathcal{D}_{1:t-1}) = (2\pi)^{-\frac{O_1}{2}}|\bm{F}_{1it}^{s,s'}|^{-\frac{1}{2}} \exp \left\{ -\frac{1}{2} (\bm{v}_{it}^{s,s'})^T [\bm{F}_{1it}^{s,s'}]^{-1} \bm{v}_{it}^{s,s'} \right\}.
\end{equation}
$f(\bm{y}_{1it}| \mathcal{D}_{1:t-1})$ in Equation (\ref{HF2}) is called the \textit{prediction error decomposition function} \citep{Schweppe1965}. The transition probabilities $Pr[S_{it}=s|S_{i,t-1}=s']$ are expressed in the same way as Equation (\ref{MS}). In evaluating them, $\bm{\eta}_{1i,t-1}$ is set to its regime-conditional filtered mean and $\bm{\eta}_{2i}$ to its Bartlett score, so the transition model conditions on point estimates of both latent predictors; the consequences of this choice are quantified in Section~\ref{sec:sim} and taken up in Section~\ref{sec:discuss}. Compared with A.3 in \cite{Chow2013}, our framework has greater flexibility because we allowed the heterogeneous transition probabilities depending on the intra-individual latent factor scores, inter-individual latent factor scores, and their interactions.

\subsection{The collapsing process (step 5)}

The approximation is based on collapsing the joint estimates $\bm{\eta}_{1it|t}^{s,s'}$ and $\bm{P}_{1it|t}^{s,s'}$ at the end of each occasion into marginal estimates $\bm{\eta}_{1it|t}^{s}$ and $\bm{P}_{1it|t}^{s}$. At each time point $t$, the Kalman filter procedures utilize only the marginal estimates: $\bm{\eta}_{1i,t-1|t-1}^{s'}$ and $\bm{P}_{1i,t-1|t-1}^{s'}$, from the previous time point. Without this procedure, computational costs would be enormous due to the necessity of executing Equations (\ref{Kalman1}) and (\ref{Kalman2}) for $2^T$ combinations and storing all those results. To circumvent the issue, \cite{Kim&Nelson1999} proposed collapsing $\bm{\eta}_{1it|t-1}^{s,s'}$ and $\bm{P}_{1it|t-1}^{s,s'}$ as follows:
\begin{align} \label{collapse}
    \bm{\eta}_{1it|t}^{s} &= \sum_{s'} W_{it}^{s,s'} \bm{\eta}_{1it|t}^{s,s'}, \\
    \bm{P}_{1it|t}^{s} &= \sum_{s'} W_{it}^{s,s'} \left[ \bm{P}_{1it|t}^{s,s'} + (\bm{\eta}_{1it|t}^{s} - \bm{\eta}_{1it|t}^{s,s'}) (\bm{\eta}_{1it|t}^{s} - \bm{\eta}_{1it|t}^{s,s'})^{T} \right], \\
    W_{it}^{s,s'} &= \frac{Pr[S_{it}=s, S_{i,t-1}=s' | \mathcal{D}_{1:t}]}{Pr[S_{it}=s | \mathcal{D}_{1:t}]}.
\end{align}
Here, $W_{it}^{s,s'}$ is referred to as the weighting factor. The computation of the weighting factor makes use of $Pr[S_{it}=s, S_{i,t-1}=s' | \mathcal{D}_{1:t}]$ and $Pr[S_{it}=s| \mathcal{D}_{1:t}]$ which are calculated in Equations (\ref{HF1}) to (\ref{HF3}).

\subsection{Marginalization of the random intercepts}\label{sec:integrated}

The person-specific random intercept residual $\bm{\zeta}_{2i}$ is constant over time, so the filter can learn about it only through the stable shift it induces in person $i$'s series. Within a single finite series, such a stable shift cannot be told apart from persistent autoregressive dynamics; formally, the augmented system is not observable. Treating $\bm{\zeta}_{2i}$ as an unknown fixed state therefore fails: the initial variance assigned to its block is an arbitrary choice that the data cannot override, and it propagates into all filtered covariances \citep[cf.][]{Asparouhov2018}. We therefore treat $\bm{\zeta}_{2i}$ as a Gaussian random effect. The state vector is augmented to
$\bm{x}_{it} = (\bm{\eta}_{1it}^{\top}, \bm{\zeta}_{2i}^{\top})^{\top}$ with transition
\begin{equation} \label{AugState}
\bm{x}_{it} = \begin{pmatrix}\bm{b}_{1s} + \bm{B}_{2s}\hat{\bm{\eta}}_{2i}\\ \bm{0}\end{pmatrix}
+ \begin{pmatrix}\bm{B}_{3is} & \bm{I}\\ \bm{0} & \bm{I}\end{pmatrix} \bm{x}_{i,t-1}
+ \begin{pmatrix}\bm{\zeta}_{1its}\\ \bm{0}\end{pmatrix},
\end{equation}
measurement $\bm{y}_{1it} = [\bm{\Lambda}_{1s}\ \bm{0}]\,\bm{x}_{it} + \bm{\epsilon}_{1its}$, zero process noise on the $\bm{\zeta}_{2i}$ block, and the proper distribution $\bm{\zeta}_{2i} \sim \mathcal{N}(\bm{0}, \bm{Q}_2)$ as its initial distribution. Because the model is linear and Gaussian conditional on the regime path, the prediction error decomposition computed by the Kalman filter is then exactly the marginal likelihood
\begin{equation} \label{MargLik}
\int p(\bm{y}_{1i,1:T} \mid \bm{\zeta}_{2i}, \bm{\theta})\,
\mathcal{N}(\bm{\zeta}_{2i}; \bm{0}, \bm{Q}_2)\, d\bm{\zeta}_{2i}:
\end{equation}
the filter never conditions on a point estimate of $\bm{\zeta}_{2i}$, but propagates its uncertainty into every innovation covariance. $\bm{Q}_2$ enters the likelihood through the initial covariance and is estimated by maximum likelihood together with all other parameters; it is identified from the cross-sectional replication over the $N$ persons, exactly as a variance component in a mixed-effects model, and its standard error is obtained in the same way as for the remaining parameters (step~6). The initial distribution of the augmented state is specified jointly, because the two blocks are dependent: $\bm{\zeta}_{2i}$ shifts the intercept of the within-level dynamics, so a person with a larger intercept residual has, at stationarity, a correspondingly shifted within-level state. Solving the stationary equations of the augmented system in regime 1 yields $\operatorname{Cov}(\bm{\eta}_{1i0}, \bm{\zeta}_{2i}) = (\bm{I}-\bm{B}_{3i1})^{-1}\bm{Q}_2$, and the stationary covariance of $\bm{\eta}_{1i0}$ equals the within-level stationary covariance of Section~\ref{sec:est:init} plus $(\bm{I}-\bm{B}_{3i1})^{-1}\bm{Q}_2(\bm{I}-\bm{B}_{3i1})^{-\top}$. No arbitrary constant remains in the initialization: the variance of the $\bm{\zeta}_{2i}$ block is not a tuning constant but the estimated $\bm{Q}_2$ itself. The marginalization is exact conditional on the regime path; the collapsing approximation of step~5 and the plug-in treatment of $\hat{\bm{\eta}}_{2i}$ (step~2) are the two approximations that remain, and their consequences are quantified in Section~\ref{sec:sim} and taken up in Section~\ref{sec:discuss}.

We refer to this estimator as the \emph{integrated} (marginal maximum likelihood) estimator: the separation of between- from within-person variation is achieved by construction, because the likelihood weighs the autoregressive and the between-person explanations of stable individual levels jointly.

\subsection{Parameter update (step 6)}
To obtain the (log-)total likelihood function $\log f(\bm{Y}_{1}|\bm{\theta})$, we sum up the log of prediction error decomposition function over time and across persons: \begin{equation} \label{sumLik}
    \log f(\bm{Y}_{1}| \bm{\theta}) = \sum_{i} \sum_{t} \log f(\bm{y}_{1it}| \mathcal{D}_{1:t-1}).
\end{equation}
We then adapt the parameters based on the optimizer of choice. The convergence criteria is also based on the total likelihood function. Let $\hat{\bm{\theta}}_{k}$ be the parameter estimate at $k$th iteration step, improvement of the total likelihood is calculated as
\begin{equation} \label{conv}
    \Delta \log f(\bm{Y}_1 | \hat{\bm{\theta}}_{k}) = \log f(\bm{Y}_1 | \hat{\bm{\theta}}_{k}) - \log f(\bm{Y}_1 | \hat{\bm{\theta}}_{k-1})
\end{equation}
which serves as the basis of convergence.

Point estimates of all
parameters $\hat{\bm{\theta}}$ can be obtained by maximizing the total log-likelihood function $\log[f(\bm{Y}_{1}|\bm{\theta})]$. The corresponding standard errors are obtained by taking the square root of the diagonal elements of the negative numerical Hessian matrix at the convergence point $\bm{I}(\hat{\bm{\theta}})$, or its approximation using the outer product of the gradient (OPG). Let $g_{i}(\bm{\theta}) = \frac{\partial}{\partial \bm{\theta}} \sum_{t} \log f(\bm{y}_{1it}| \mathcal{D}_{1:t-1})$,
\begin{equation}
    \bm{I}_{\text{OPG}}(\bm{\theta}) = \sum_{i=1}^{N} g_{i}(\bm{\theta}) g_{i}(\bm{\theta})^{T}.
\end{equation}
The information matrix equality $\bm{I}(\bm{\theta}) = \bm{I}_{\text{OPG}}(\bm{\theta})$ holds given a correctly specified model and standard regularity assumptions \citep[see][]{White1982}.

\subsection{Missing values}

Missing values often occur in ILD in psychology. \cite{Harvey1990} successfully accommodated missing values in the Kalman filter without regime switching. When the time-dependent observed variables $\bm{y}_{1it}$ contain missing values,
the one-step-ahead prediction error in step 3.2 cannot be calculated. The consequences are that the likelihood function cannot be obtained and the updated estimates in step 3.3 cannot be determined. As a result, updated estimates are replaced by predicted estimates. This also applies for the Kalman filter.

In the case of our extended Kim filter, missing values not only affect the Kalman filter, but also the extended Hamilton filter. To overcome this additional issue, one can use the method proposed by \cite{Hamaker2012}. At each time point, step 3.2 and 3.3 of the Kalman filter (Equation \ref{Kalman3} to \ref{Kalman7}), the step 4.2 of the extended Hamilton filter (Equation \ref{HF2} to \ref{HF4}), and the likelihood calculation (Equation \ref{lik}) will be skipped for subjects with missing values. Instead, estimations in the update step $\left( \bm{\eta}_{1it|t}^{s,s'}, \bm{P}_{1it|t}^{s,s'}, Pr[S_{it}=s, S_{i,t-1}=s' | \mathcal{D}_{1:t}] \right)$ are substituted by the estimations in the prediction step $\left( \bm{\eta}_{1it|t-1}^{s,s'}, \bm{P}_{1it|t-1}^{s,s'}, Pr[S_{it}=s, S_{i,t-1}=s' | \mathcal{D}_{1:t-1}] \right)$.

\section{Empirical Study}\label{sec:emp}

The empirical context of this study is introduced in Section 2. Here we describe the estimation procedure, model assumptions, and forecasting results specific to this application.

\subsection{Forecast implementation}\label{sec:emp:impl}

Prior to the parameter estimation, seven assumptions are made.
First, large fixed intercepts are assumed in the second regime (i.e, $\bm{b}_{11} < \bm{b}_{12} $). This is provided so that the second latent state can be considered to be related to an intention to dropout \citep[same assumption is made in][]{Kelava2022}.
Second, in addition to model identification constraints, factor loadings ($\bm{\Lambda}_{1s}, \bm{\Lambda}_{2}$) are assumed to be class-invariant and sparse with no cross-loadings, in order to ensure measurement invariance (MI).
Third, within- and between-level residual covariance matrix in the measurement model ($\bm{R}_{1s},\bm{R}_{2}$) and the structural model ($\bm{Q}_{1s}, \bm{Q}_{2}$) are assumed to be diagonal and class-invariant in order to assume local independence and reduce model complexity.
Fourth, autoregressive coefficient matrix $\bm{B}_{3is}$ is assumed to be stationary and diagonal \citep[i.e., AR(1) model is used; cf., Footnote 2 of][]{Kelava2022}. Stationarity was imposed for practical reasons because violations led to numerical overflow in the covariance recursion; for the diagonal AR(1) specification, it requires $|(\bm{B}_{3is})_{jj}| < 1$ for $j=1,\ldots,U_1$ and $s \in \{1,2\}$.

Fifth, the (transition) probability to return to $S_{it}=1$ after $S_{i,t-1}=2$ was fixed to a small value ($P_{12} = 10^{-12}$) for all individuals $i$ and time $t$ for model simplicity and interpretation.\footnote{The same assumption was made in the original study \citep{Kelava2022}.} According to \textcite{Heublein2014}, dropout is understood as a thoughtful decision rather than a spontaneous decision, and returning is rare.
Sixth, the interaction effects ($\bm{\gamma}_{4}$) in the Markov-switching model are restricted to zero with respect to the motivational variables (\textit{content not important}, \textit{not understanding}, \textit{positive affect}, \textit{negative affect}), following \citet{Kelava2022}; interaction effects are therefore estimated only with respect to the self-regulatory variables (\textit{cost}, \textit{afraid to fail}, \textit{stress}).
Seventh, if a dropout is observed for a student $i$ at time $t$, the regime $S_{it}$ becomes an observed value i.e., $S_{it}, \ldots, S_{iT}=2$. This condition is to identify the second regime as \textit{an intention to dropout}.

Unless otherwise noted, the following initialization conventions apply to both the empirical and simulation studies. All individuals are assumed to start in regime~1: the initial regime probability is set to $\Pr[S_{i0}=1|\mathcal{D}_{1:0}]=1$ \citep[following][]{Kelava2022}, reflecting the substantive assumption that no participant has yet formed a dropout intention at the start of the semester. Because all probability mass begins in regime~1, the natural model-implied initial distribution for the state is the regime-1 \emph{conditional} stationary distribution, with the stationary mean and covariance given in closed form in Section~\ref{sec:est:init}, well defined because the regime-1 dynamics are stationary ($|(\bm{B}_{3i1})_{jj}|<1$; Assumption~4). The regime-\emph{averaged} alternatives would also fail on formal grounds: because regime~2 is treated as (nearly) absorbing ($P_{12} \approx 0$; Assumption~5), the regime-averaged stationary distribution degenerates onto the absorbing regime, and the ergodic distribution sometimes used for the initial regime probabilities likewise places almost all mass on regime~2. Under the integrated estimator, the initial distribution is the joint stationary distribution of $(\bm{\eta}_{1i0}, \bm{\zeta}_{2i})$: the $\bm{\zeta}_{2i}$ block is initialized at $\mathcal{N}(\bm{0}, \bm{Q}_2)$ and the cross-covariance between the blocks is $(\bm{I}-\bm{B}_{3i1})^{-1}\bm{Q}_2$ (Section~\ref{sec:integrated}). Initial parameter values were drawn from normal distributions centered at the posterior means reported in \textcite{Kelava2022}, with standard deviations equal to the corresponding posterior standard deviations. Five independent optimization runs from different random starting values (seeds) were performed, and the highest-likelihood solution was retained. For the empirical study, the five runs attained log-likelihoods within $0.002$ of one another (per-observation log-likelihood between $-17.6986$ and $-17.7002$), with three of the five runs locating the same best optimum. Across these near-equivalent optima, all measurement parameters, the between-person effects $\bm{B}_{21}$ and $\bm{B}_{22}$, the autoregressive diagonals (with the single exception discussed in Section~\ref{sec:emp:sm}), and the variance components $\bm{Q}_1$ and $\bm{Q}_2$ were stable (across-run ranges at most $0.13$), whereas the Markov-switching transition coefficients varied across the optima by amounts comparable to their own magnitude. We treat the across-run variability of the transition coefficients as direct evidence of their weak identification (Section~\ref{sec:discuss}); the regime predictions were robust across the optima (the sets of students classified in the dropout-intention regime at $t = 50$ agree for at least $113$ of the $119$ students, $94.9\%$, between any pair of optima, and coincide exactly among the three runs attaining the best optimum).

The estimation procedure was implemented in Python \citep{Rossum2009}. The model estimation and forecasting were implemented according to Algorithm 1; the optimizer settings and the convergence criterion are given in Appendix~\ref{app:empirical}. The first half of the measurements was used to estimate the parameters, and the second half was used to evaluate the out-of-sample forecasting performance.

\subsection{Forecasting results}

\subsubsection{Markov-switching parameters}
Parameter estimates are found in Table \ref{tab:emp_params_ms} (see appendix). The baseline transition intercept was $\gamma_{1} = 1.53$ (SE $= 0.63$), consistent with the posterior mean of $1.48$ reported by \textcite{Kelava2022}; this coheres with the simulation finding that $\gamma_1$ is the one well-identified transition coefficient (Section~\ref{sec:sim}). The main between-individual effect was $\gamma_{2} = -0.11$ (SE $= 1.51$); it is negative in all five runs but varies across the near-equivalent optima ($-0.94$ to $-0.11$), so we do not interpret its magnitude. The main intra-individual effects ($\bm{\gamma}_{3}$, one coefficient per within-individual indicator) and the interaction effects ($\bm{\gamma}_{4}$, free only at the three self-regulatory indicators; Assumption~6) were freely estimated; because these are vector-valued, we report their individual components in Table~\ref{tab:emp_params_ms} rather than in the text. Their estimates were moderate in magnitude (all within $\pm 1.7$), but their standard errors ($1.4$ to $3.8$) exceed the estimates throughout, none is statistically distinguishable from zero, and the estimates vary across the near-equivalent optima. We therefore report these coefficients with their standard errors and do not interpret their sign or magnitude; the across-run variability at essentially identical likelihood is itself the clearest empirical evidence of their weak identification (see Section~\ref{sec:discuss}).

\subsubsection{Structural model parameters}\label{sec:emp:sm}
\textbf{No intention to dropout $(s=1)$}.
Parameter estimates are found in Table \ref{tab:emp_params_sm1} (see appendix). The baseline effect ($\bm{B}_{21}$) of \textit{cognitive skills} on the intercept was negative or near zero across the 7 within-scales ($0.00$ to $-0.26$). This means that persons with higher cognitive skills had lower scores on the within-scales such as \textit{cost}. The autoregressive coefficients (diagonal elements of $\bm{B}_{31}$) indicate moderate temporal dependency: for six of the seven within-scales they ranged from $0.23$ to $0.77$; stable individual differences in level are carried by the between-person components ($\bm{B}_{21}\hat{\bm{\eta}}_{2i}$ and the marginalized $\bm{\zeta}_{2i}$) rather than by the autoregression. The one exception is \textit{afraid to fail}, whose process-noise variance is estimated at essentially zero ($\bm{Q}_{1,3} = 0.000$ in the reported run, at most $0.011$ across runs): the within-level dynamics of this factor are degenerate, so its autoregressive coefficient is not identified ($-0.12$, SE $= 0.50$; ranging from $-0.13$ to $+0.19$ across runs). The effect of cognitive skills on within-level factor dynamics ($\bm{B}_{41}$) was small and largely overlapping zero across all scales.

\textbf{Intention to dropout $(s=2)$}.
Parameter estimates are found in Table \ref{tab:emp_params_sm2} (see appendix). Like the first state, the baseline effect ($\bm{B}_{22}$) of \textit{cognitive skills} on the intercept was negative or near zero for all 7 within-scales. The autoregressive coefficients (diagonal elements of $\bm{B}_{32}$) were positive and moderate ($0.43$ to $0.84$), and larger than the corresponding first-state coefficients for six of the seven scales (and essentially equal for the seventh), so the dropout-intention regime showed somewhat stronger persistence. The state-2 specific parameters carried larger standard errors, indicating limited precision in estimating dynamics specific to the dropout-intention regime.

\subsubsection{Variance estimation}
Parameter estimates are found in Table \ref{tab:emp_params_sm3} (see appendix). The within-level process noise variances (diagonal elements of $\bm{Q}_1$) were relatively larger for \textit{no positive affect} and \textit{negative affect}, implying greater fluctuation in these affective states across time; the variance for \textit{afraid to fail} is estimated at essentially zero, the degenerate case noted above. The between-level variances (diagonal elements of $\bm{Q}_2$), estimated by maximum likelihood within the filter (Section~\ref{sec:integrated}), ranged from $0.00$ (\textit{content not important}) to $0.19$ (\textit{negative affect}). These are nonnegligible for most scales, implying that stable inter-individual differences are not exhausted by baseline differences in cognitive skills.

\subsubsection{Measurement model parameters}
Parameter estimates for the measurement model are found in Tables \ref{tab:emp_params_mm1} and \ref{tab:emp_params_mm2} (see appendix). The within-level factor loadings ($\bm{\Lambda}_{1,\text{free}}$) were all positive and substantial ($0.90$ to $1.43$), indicating that the observed items load strongly on their respective latent factors, and were the most stable quantities across the five runs (across-run ranges below $0.02$). The within-level residual variances ($\bm{R}_1$) were small relative to the within-level factor variances, suggesting adequate measurement quality at the within level. At the between level, the cognitive-skills factor had an estimated variance of $\bm{P}_2 \approx 0.87$, against measurement-error variances of $\bm{R}_2 \approx 0.32$ and $1.81$ for its two indicators, corresponding to indicator reliabilities of about $.73$ and $.32$. Measurement error in the observed between-person scores is therefore substantial: substituting $\bm{y}_{2i}$ directly for the latent factor would treat these scores as error-free and attenuate the between-level regression coefficients accordingly, which is the empirical rationale for the latent-variable treatment of cognitive skills. One caveat applies: the standard errors for $\bm{R}_2$ were very large, because under the present algorithm $\bm{\Lambda}_2$ and $\bm{R}_2$ are identified only through the plug-in factor scores rather than from the between-person covariance of $\bm{y}_{2i}$ (Section~\ref{sec:integrated}), so the reliability figures above are themselves imprecise; the full marginalization of $\bm{\eta}_{2i}$ discussed in Section~\ref{sec:discuss} would identify them from their own likelihood contribution.

\subsubsection{Regime probabilities}\label{sec:emp:regimeprob}
Compared with the 45 persons (37.8\%) who had actually dropped out, the model classified 111 individuals (93.3\%) in the dropout-intention regime ($\Pr[S_{it}=2] > 0.5$) at the final time point ($t = 50$). Read as a prediction of actual dropout, the default 0.5 cutoff is thus markedly liberal: it flags nearly every student, and roughly two and a half times as many as eventually drop out. Three considerations frame this result. First, the model predicts a latent intention to dropout, whereas the observed outcome is actual dropout: under the Rubicon model of action phases \citep{Achtziger2018}, an intention precedes, and need not culminate in, dropout, so students flagged as having a dropout intention who did not (observably) drop out are not necessarily misclassified. Second, the absorbing structure of regime~2 (Assumption~5) means that any crossing of the 0.5 threshold, however early, persists to the end of the series, so the fraction of flagged students can only grow. Third, a like-for-like evaluation of regime recovery, in which the true regime is known, is provided by the simulation study, where the same estimator attains specificity of $.87$--$.94$ (Table~\ref{tab:sim_metrics}); the apparent over-flagging relative to \emph{observed dropout} therefore reflects the intention/behavior gap and the cutoff, not an inability of the filter to discriminate regimes.

We therefore focus on how early the model detects the students who do drop out. Because the Hamilton filter enforces $S_{it}=2$ once $\text{DO}_{it}=1$ is observed (Assumption~7), regime probabilities reach 1.0 at the event itself; we therefore distinguish \textit{genuine forecasts} (crossing the 0.5 threshold at least one time point before dropout) from \textit{same-time detections}. All 45 actual dropouts were flagged at or before the event: 42 ($93.3$\%) were genuine forecasts, with a mean lead time of 34.3 time points (median $= 42.5$), and three were same-time detections. The flags arise very early (mean first crossing of the 0.5 threshold at $t \approx 7$ among the 45 eventual dropouts), whereas the dropout events occur much later (mean $t = 37.4$, SD $= 15.9$), clustering at the end of the semester (25 of the 45 at the final time point, $t = 50$); part of the long lead therefore reflects that the behavioral event is recorded late, not only that the intention forms early. At-risk students are thus identifiable early, largely from baseline and early observations rather than from a transition detected shortly before dropout, which is the pattern expected under the Rubicon model \citep{Achtziger2018}: an intention is formed well before, and need not immediately precede, the behavioral act. The early onset also partly reflects the informative initial conditions and the absorbing structure of regime~2, and it is consistent with the weak identification of the within-person transition predictors.

\begin{figure}
    \centering    \includegraphics[width=\linewidth]{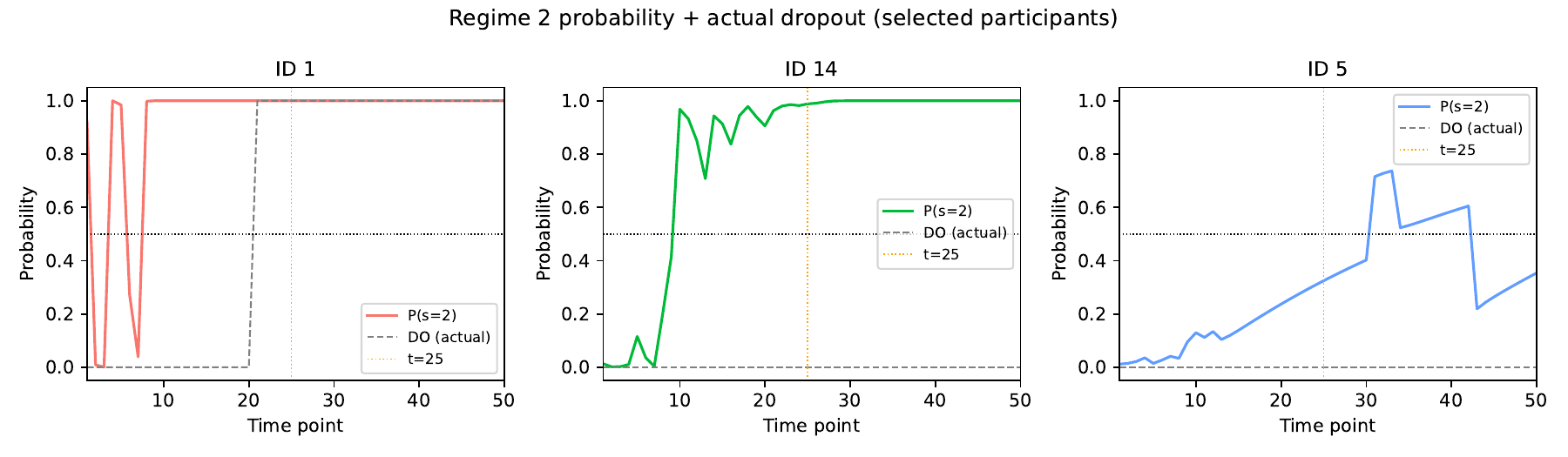}
    \caption{Comparison of predicted regime probability (marginalized) and true dropout occurrence for selected individuals. The solid lines represent predicted values, while the dashed line represents true values. The dotted line at \( y=0.5 \) serves as a cutoff point} \label{fig:emp_do_predicted}
\end{figure}

Figure \ref{fig:emp_do_predicted} illustrates the trajectories of the regime probabilities corresponding to \textit{intention to dropout} for three individuals compared with their actual dropout status. The left panel shows that student (\#1) signaled an intention to dropout within the first few time points and dropped out at $t \approx 21$. The middle panel shows that student (\#14) maintained a low dropout-intention probability only briefly before rising sharply to a sustained high value around $t \approx 10$, but did not (observably) drop out. The right panel shows that student (\#5) drifted upward only slowly, exceeding the 0.5 threshold only modestly between $t = 31$ and $t = 42$ (peaking at $.74$) before falling back below it and ending the series there; this student also did not drop out, and the panel illustrates what a borderline trajectory looks like under the default cutoff.

These results sharpen the practical guidance for early-warning use. With near-universal flagging at the default cutoff, $\Pr[S_{it}=2] > 0.5$ is not, by itself, an actionable intervention trigger in this application: an institution cannot usefully intervene on 93\% of a cohort. Nor is raising the cutoff sufficient by itself: the probability trajectories saturate near their upper bound, so that at $t = 50$, 103 of the 111 flagged students exceed $.99$ (97 exceed $.999$), and even a cutoff of $.99$ still flags $87\%$ of the cohort. The discriminating information thus lies less in the final level of a trajectory than in when and how firmly it rises (Figure~\ref{fig:emp_do_predicted}); richer functionals than a fixed cutoff at a fixed time, such as the time at which the trajectory first crosses a threshold or the steepness of its rise, are therefore the natural candidates for ranking students by urgency. The tradeoff across all possible cutoffs can be visualized as a Receiver Operating Characteristic (ROC) curve, providing practitioners with a principled tool for selecting a threshold that balances the costs of missed detection and unnecessary intervention in their specific context; the default cutoff is simply one operating point on this curve, and any operating point with fewer false alarms is available at the price of delayed or missed detections. Constructing such a curve requires outcome labels: in the simulation study (Section~\ref{sec:sim}), where true regime membership is known, predicted and true regimes are directly comparable, and the reported sensitivity and specificity are operating points of exactly such curves; in the empirical application, the only observed label, actual dropout, measures a different construct from the dropout intention the model tracks, so an empirical ROC curve compares intention forecasts with a downstream outcome and should be read with that distinction in mind.

\subsubsection{Intra-individual latent factor scores}
\begin{figure}
    \centering
    \includegraphics[width=0.8\linewidth]{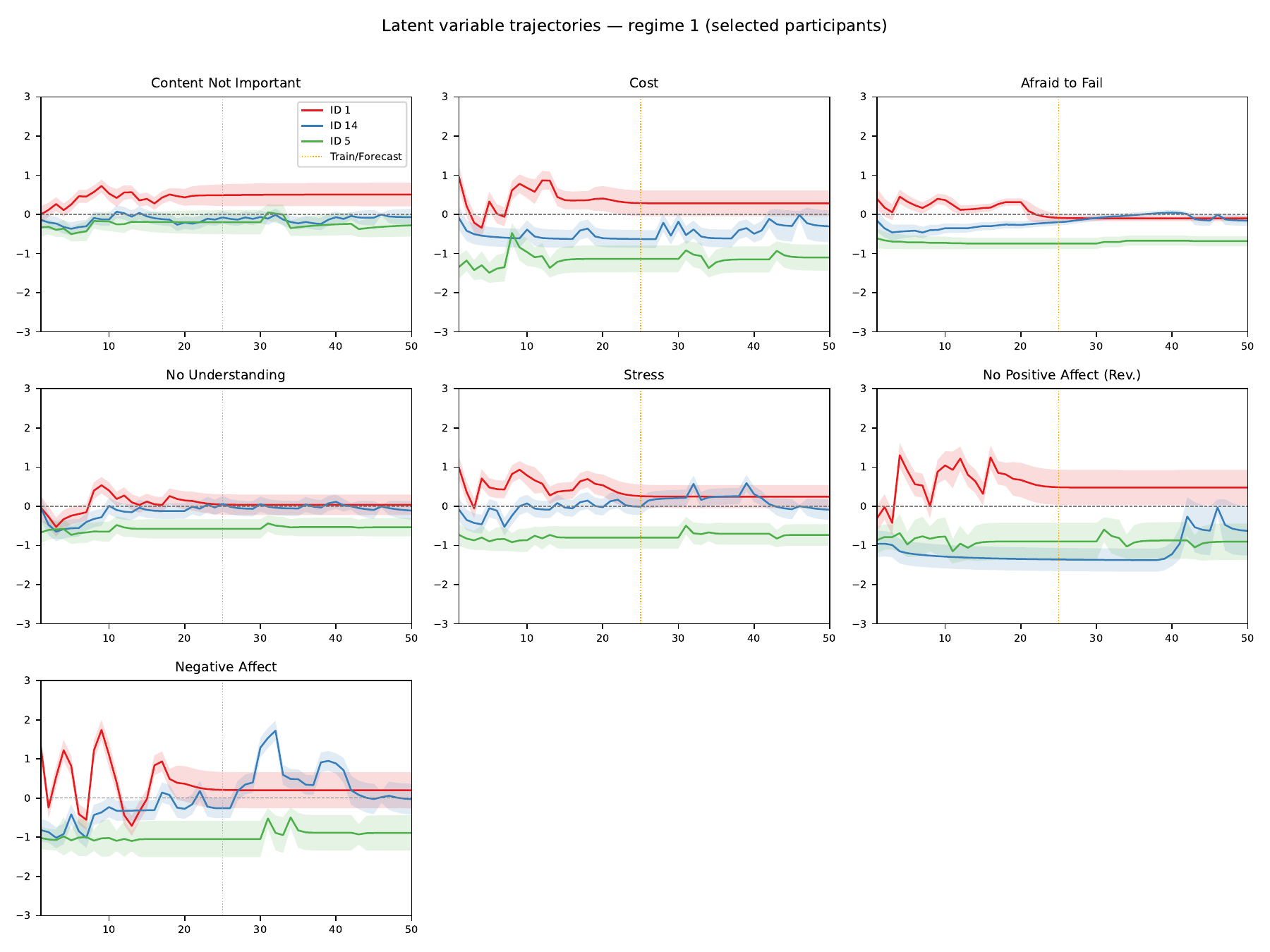}
    \caption{Comparison of Factor Scores for Selected Individuals. Each panel represents a unique factor. The solid colored lines indicate factor scores for each individual, while the shaded regions represent variance bands (\( \pm 1 \) standard deviation)}
    \label{fig:emp_latent}
\end{figure}

Figure~\ref{fig:emp_latent} illustrates the intra-individual latent factor trajectories for the same three students. The student who dropped out (\#1) exhibited consistently higher scores in \textit{content not important} and \textit{cost}, and consistently reduced positive affect, compared with the two students who did not drop out (\#14 and \#5); \textit{stress} and \textit{negative affect} were likewise markedly elevated over roughly the first 15 time points, although student \#14 reaches comparable levels later in the series (after \#1's dropout at $t \approx 21$, the plotted trajectory is a model extrapolation with no new observations). Student \#5 has among the lowest scores throughout, and student \#14 does not display the elevated latent profile of the actual dropout; yet student \#14 is assigned a sustained high dropout-intention probability from $t = 10$ onward, well before this student's later elevations in stress and negative affect, and student \#5 hovers near the cutoff over the second half of the series (see Figure~\ref{fig:emp_do_predicted}), illustrating that the predicted intention does not map one-to-one onto the measured risk factors and reinforcing the cutoff discussion above.

\section{Simulation Studies}\label{sec:sim}
In this section, we present a simulation study that investigates the performance of the proposed approach given complete datasets. Regime prediction, prediction of the dynamic latent variables, and parameter retrieval are evaluated with a population model that is built upon the characteristics of the empirical example in the previous section. 

\subsection{Data generating process}
Data were generated using the parameter estimates reported by \citet{Kelava2022} as the population (true) values, providing an external, published calibration for the data-generating process. In the simulation, we used only 2 dynamic latent factors (\textit{cost, afraid to fail}). This reduced model complexity was necessary to keep computation time feasible for a simulation study. Two parameters governing the regime process were set by us rather than taken from \citet{Kelava2022}. First, the switch-back probability $P_{12}$ is fixed at $10^{-12}$ rather than the estimated $0.097$, which keeps the data-generating process consistent with Assumption~5 (Section~\ref{sec:emp:impl}), under which the estimator treats regime~2 as nearly absorbing; generating data with a reversible transition while estimating under a near-absorbing constraint would render the estimator misspecified, so that parameter recovery could no longer be separated from the effect of that misspecification. The true value of $P_{12}$ in Table~\ref{tab:sim_params_ms} therefore reads $0.00$. Second, the transition intercept is set to $\gamma_1 = 3.50$ rather than the reported $1.48$: under the near-absorbing regime~2, the reported value would leave almost no regime-1 observations after the first few occasions and render the classification metrics uninformative, whereas at $\gamma_1 = 3.50$ (a per-time-point switching probability of about $3\%$) both regimes are represented throughout the design. All measurement and structural parameters are those of \citet{Kelava2022}.
We adopted a $2 \times 2$ factorial design crossing sample size $N \in \{50, 100\}$ and estimation length $T_{\text{est}} \in \{25, 50\}$, yielding four conditions: $\mathcal{D}_{50,25}$, $\mathcal{D}_{100,25}$, $\mathcal{D}_{50,50}$, and $\mathcal{D}_{100,50}$. This design mirrors that of \cite{Kelava2022}, who crossed $N \in \{25, 50\}$ with $N_t \in \{25, 50\}$ and a fixed 10-point forecast window ($N_t^{+}=10$); our larger sample sizes were chosen so that the upper level ($N=100$) approximates the sample size of the empirical application ($N=119$). $T=60$ time points were generated for all replications to ensure a consistent data-generating process across conditions. The first 50 time points follow the regime-switching dynamics implied by these population values; the remaining 10 time points continue under the same parameter values. For the $T_{\text{est}}=25$ conditions, only $t=1,\ldots,35$ are analysed (25 estimation + 10 forecast); for the $T_{\text{est}}=50$ conditions, all 60 time points are used. The simulation was replicated $R=100$ times per condition ($400$ total).

\subsection{Forecast implementation and evaluation}

Estimation and forecasting follow Algorithm 1, with five independent initializations per replication. Initialization of state moments, regime probabilities, and parameter starting values follows the conventions described in Section~\ref{sec:emp}, and the optimizer settings and the convergence criterion are given in Appendix~\ref{app:empirical}. Depending on the condition, the first 25 or 50 time points were used to estimate the parameters, and the subsequent 10 time points ($t=26,\ldots,35$ for $\mathcal{D}_{50,25}$ and $\mathcal{D}_{100,25}$; $t=51,\ldots,60$ for $\mathcal{D}_{50,50}$ and $\mathcal{D}_{100,50}$) were used to evaluate the out-of-sample forecasting performance, consistent with the $N_t^{+}=10$ forecast window used in \cite{Kelava2022}.

For evaluation, first, we investigated the \textit{sensitivity} and \textit{specificity} for the prediction of the underlying regime memberships. We distinguished the performance within the observed data ($t=1,\ldots,25$ or $t=1,\ldots,50$) and the forecast data ($t=26,\ldots,35$ or $t=51,\ldots,60$, depending on the condition). Second, a score function is used to evaluate the forecast of the dynamic latent factors by
\begin{equation}
    \delta_{t} = \frac{1}{2N} \sum_{i} \sum_{j} (\hat{\eta}_{itj} - \eta_{itj})^{2}, \hspace{10pt} i=1,\ldots,N,  \hspace{5pt} j=1,2
\end{equation}
for each forecast time point \citep[cf.,][]{Gneiting2011}. Third, our parameter estimation was evaluated in terms of the bias compared to the true values, standard deviation of estimates across replications, and RMSE. Convergence was $99$ of $100$ replications in every condition; a single generated dataset produced a singular innovation covariance under all five starts and was excluded.

\subsection{Forecasting results}

\textit{Parameter estimation} Tables \ref{tab:sim_params_ms}, \ref{tab:sim_params_sm}, and \ref{tab:sim_params_mm} summarize the results (see appendix). The within-level measurement residuals ($\bm{R}_1$) were recovered essentially without error (bias $\approx 0.00$, RMSE $\leq 0.02$, coverage at $1.00$), and the within-level process-noise variances ($\bm{Q}_1$) showed only a negligible upward bias ($\leq 0.01$). The autoregressive coefficients $\mathrm{diag}(\bm{B}_{3s})$, with true values $0.15$--$0.69$, were recovered with biases between $-0.02$ and $+0.10$ (coverage $.40$--$.99$), and the random-intercept variance $\bm{Q}_2$ was estimated with a small upward bias ($+0.02$ to $+0.05$) and valid standard errors from the same OPG computation as the remaining parameters; the marginalization thus separates autoregressive persistence from stable person-specific levels. The baseline transition rate $\gamma_1$ (true value $3.50$) was well identified, with bias between $+0.08$ and $+0.21$, RMSE between $0.26$ and $0.54$, coverage between $.79$ and $.97$, and power at $1.00$ in every condition. (The between-person effect $\gamma_2$ is discussed below, together with the other coefficients that load on the plug-in factor score, because its bias has a different source.) The within-person transition effects $\bm{\gamma}_3$ and $\bm{\gamma}_4$ remained weakly identified, but they fail in a predictable way. Their estimates are shrunk toward zero rather than distorted in an arbitrary direction (biases within about $0.6$ in absolute value): the transition model sees the filtered within-person state rather than the true latent state, and this estimation error attenuates the coefficients, just as measurement error in a regressor would. The standard errors report this uncertainty honestly, with coverage of $.73$--$.99$, so power is correspondingly low ($.00$--$.40$, exceeding $.20$ only in the largest design; Table~\ref{tab:sim_power}). The dominant limitation is imprecision: because regime~2 is nearly absorbing, each person contributes no more than a single transition, and under the population values a movement of one standard deviation in $\bm{\eta}_{1i,t-1}$ shifts the per-occasion switching probability by less than one percentage point, so the data contain few informative events for these coefficients; consistently, the intervals are wide but their coverage is close to nominal.

A well-understood set of biases remains, and it is attributable to the one plug-in still present in the algorithm: the Bartlett factor score $\hat{\bm{\eta}}_{2i}$ (step~2). The between-person effects were attenuated toward zero (e.g., $\bm{b}_{21}$ with true value $-0.87$ was estimated around $-0.26$, bias $+0.61$; $\gamma_2$ with true value $-1.17$ was estimated near zero, at $-0.12$ to $-0.16$), and the between-person factor variance $P_2$ (true value $0.20$) was inflated to $0.66$--$0.71$. Coverage for $\gamma_2$ dropped as low as $.03$: the estimator is precise about the attenuated value. For the linear channel the attenuation is quantitatively consistent with classical measurement-error theory: the Bartlett score has error variance $(\bm{\Lambda}_2^{\top} \bm{R}_2^{-1} \bm{\Lambda}_2)^{-1} \approx 0.386$ under the population values, so with $P_2 = 0.20$ the implied attenuation factor is $0.20/0.586 \approx 0.34$, predicting $\hat{\bm{b}}_{21,1} \approx -0.30$ (observed $-0.26$) and $\hat{P}_2 \approx 0.59$ (observed $0.66$--$0.71$). The shrinkage of $\gamma_2$ is stronger than this linear factor predicts, as expected when the error-contaminated score enters a nonlinear (logistic) channel together with its interaction term. The residual biases are therefore the plug-in measurement error of $\hat{\bm{\eta}}_{2i}$, not the separation of between- from within-person variation. The between-level residual variances $\bm{R}_2$ remain imprecisely estimated for the same reason: they are identified only through the plug-in factor scores, as their covariance is not itself modeled.

Empirical power (Table~\ref{tab:sim_power}), defined as the proportion of (converged) replications in which the 95\% confidence interval excludes zero, admits a clean reading, because the estimates it refers to are essentially unbiased for the well-identified parameters. Power for the baseline transition rate $\gamma_1$ was $1.00$ throughout. Power for the autoregressive coefficients increased with both design factors and reflects genuine detection (e.g., $\mathrm{diag}(\bm{B}_{31})_2$, true value $0.15$, from $.10$ in $\mathcal{D}_{50,25}$ to $.96$ in $\mathcal{D}_{100,50}$). Power for the attenuated between-person effects $\bm{b}_{21}$ and $\bm{b}_{22}$ rose sharply with $N$ (e.g., $\bm{b}_{22,1}$ from $.20$ to $.97$), which should be read together with the attenuation above: the tests detect the (shrunken) plug-in effect. Power was low for $\bm{\gamma}_2$, $\bm{\gamma}_3$, and $\bm{\gamma}_4$ ($\leq .40$, and $\leq .19$ outside the largest design) and negligible for the variance components $\bm{Q}_1$ and $\bm{R}_2$ ($\approx 0$), the latter for the distinct reasons given above.

\textit{Regime membership}
Table \ref{tab:sim_metrics} summarizes the performance of regime-switching predictions. Sensitivity ranged from $.69$ to $.82$ within the estimation window and from $.84$ to $.91$ over the forecast window; specificity ranged from $.91$ to $.94$ and from $.87$ to $.92$, respectively. Both components are thus high simultaneously in every condition and window, so the regime predictions are highly informative throughout.

We report sensitivity and specificity, following \citet{Kelava2022} and \citet{Chow2013}. Because regime~2 cannot be left once entered ($P_{12} \approx 0$; Assumption~5), it accumulates over the series, comprising roughly $29\%$ to $46\%$ of person-time points within the estimation window and $55\%$ to $75\%$ within the forecast window. Sensitivity is higher over the forecast window than within the estimation window, because most forecast-window regime-2 time points belong to persons who switched during the estimation window and whose (absorbing) regime-2 status is easy to carry forward; specificity is somewhat lower over the forecast window, because regime-1 points become scarcer as the series proceeds and any false positive persists for the remainder of the series. The design factors had modest effects: a longer estimation window raised sensitivity (from $.69$--$.71$ to $.81$--$.82$ within the estimation window), while specificity was stable across conditions.

\begin{table}[H]
    \centering
    \caption{Sensitivity and specificity for regime prediction across the four simulation conditions. ``Observed'' refers to the estimation interval and ``Forecast'' to the 10-point forecast window}
    \label{tab:sim_metrics}
    \begin{tabular}{@{}lll l ll@{}}              \toprule
        \textbf{Data} & \multicolumn{2}{l}{\textbf{Sensitivity}} & & \multicolumn{2}{l}{\textbf{Specificity}} \\ \cline{2-3} \cline{5-6}
        & Observed & Forecast & & Observed & Forecast \\
        \midrule
        $\mathcal{D}_{50,25}$
        & .71 & .84 & & .91 & .87 \\
        $\mathcal{D}_{100,25}$
        & .69 & .85 & & .92 & .88 \\
        $\mathcal{D}_{50,50}$
        & .82 & .91 & & .92 & .89 \\
        $\mathcal{D}_{100,50}$
        & .81 & .90 & & .94 & .92 \\
        \bottomrule
    \end{tabular}
\end{table}

\begin{figure}[H]
    \centering
    \includegraphics[width=0.85\linewidth]{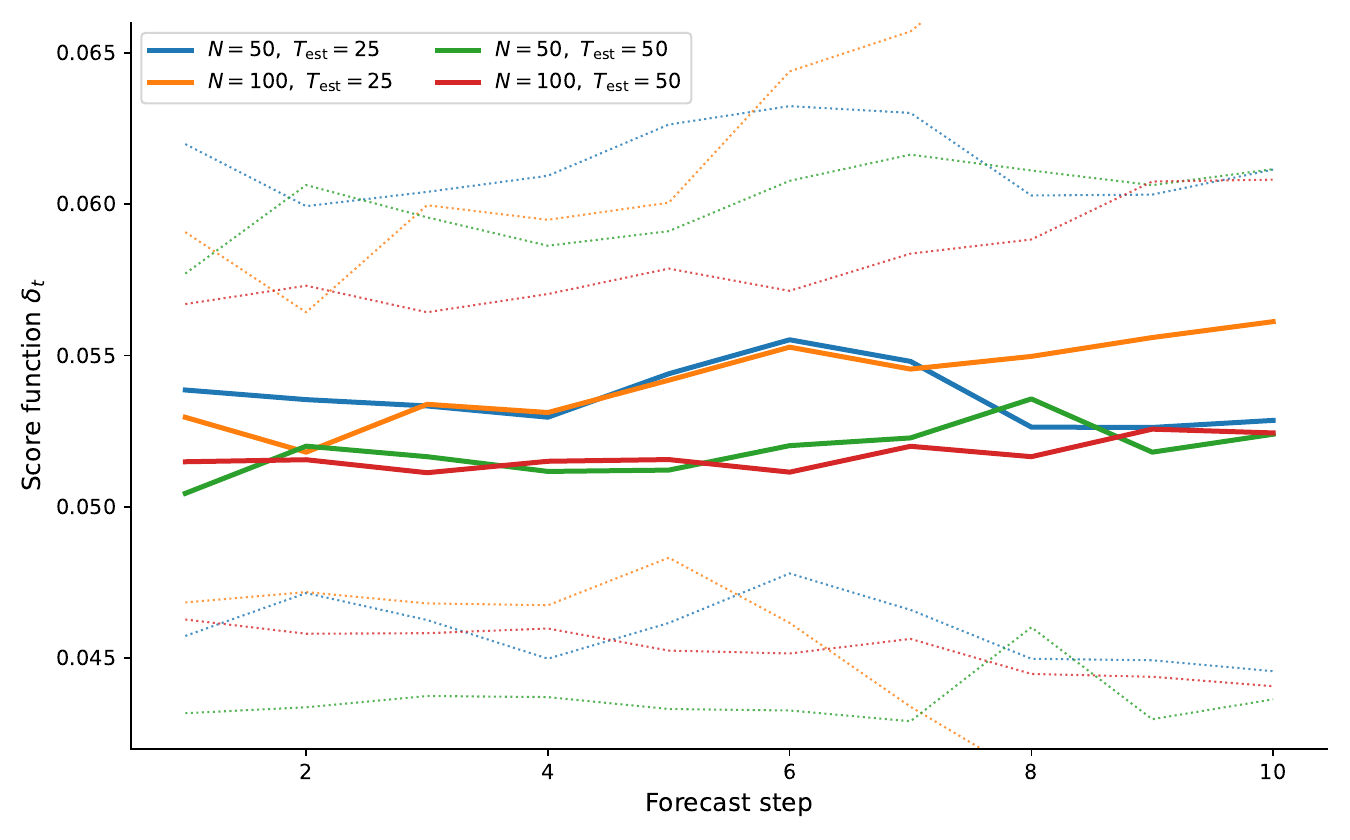}
    \caption{Score function $\delta_t$ across the 10-point forecast window under the four simulation conditions ($2 \times 2$ factorial design: $N \in \{50, 100\}$ and $T_{\text{est}} \in \{25, 50\}$), under the integrated estimator. Lines show condition means; dotted lines indicate $\pm 1$ SD across replications. The vertical axis is truncated for legibility; the $+1$ SD line for $N = 100$, $T_{\text{est}} = 25$ continues beyond the upper limit, reaching $0.077$ at the final step}
    \label{fig:sim_score}
\end{figure}

\textit{Dynamic latent variables}
The score function $\delta_t$ measures the mean squared distance between predicted and true factor scores at each forecast time point. Figure \ref{fig:sim_score} presents the score function trajectories across the forecast interval for all four conditions. Under the integrated estimator the four conditions form a narrow band, with the $T_{\text{est}}=50$ conditions slightly below the $T_{\text{est}}=25$ conditions and an essentially flat profile across the 10-step horizon; only $\mathcal{D}_{100,25}$ shows a mild upward drift toward the end of the horizon. The effect of sample size $N$ was small throughout.

\subsection{Comparison with a restricted transition model}
The transition equation in Equation~\ref{MS} nests simpler specifications as special cases (Section~\ref{sec:models}). To assess empirically what is gained by admitting the within-person latent state $\bm{\eta}_{1i,t-1}$ as a predictor of regime change, we re-ran the full design with the estimator constrained to $\bm{\gamma}_{s'3} = \bm{\gamma}_{s'4} = \bm{0}$, so that transition probabilities depend only on the intercept and the between-person factor. The data-generating process was left unchanged, so the restricted model is misspecified by construction; this is the comparison of interest, because it quantifies the cost of omitting a within-person predictor that is in fact operative. All other aspects of the design, including the number of replications and the estimation settings, were identical.

Convergence was identical ($99$ of $100$ replications per condition). The structural and measurement parameters were unaffected (e.g., $\mathrm{diag}(\bm{B}_{31})_1$ bias $+0.01$ to $+0.04$ under both specifications), and the recovery of the baseline transition rate $\gamma_1$ was essentially identical (bias $+0.07$ to $+0.23$ restricted, against $+0.08$ to $+0.21$ for the full model). Classification was also essentially unchanged: across the four conditions and both windows, sensitivity and specificity each differed between the restricted and the full model by at most $.01$, so neither specification was meaningfully better.

At these population values, then, the within-person transition predictors carry little usable information for classification: each person contributes no more than a single transition, so including $\bm{\eta}_{1i,t-1}$ neither helps nor hurts the regime predictions, and, equally, nothing is paid for including it, in estimation or in classification. Users whose interest is regime classification and forecasting therefore lose nothing with the restricted specification. The full specification remains the vehicle for substantive questions about within-person transition effects, where the quantities of interest are the signs and magnitudes of the coefficients. For that purpose the estimator behaves honestly, if in a data-hungry way: the estimates are moderately shrunk toward zero, the near-nominal coverage means that the reported intervals faithfully convey how strongly the data constrain sign and magnitude, and this precision grows with the design (e.g., power for $\bm{\gamma}_{3,1}$ rising to $.40$ in $\mathcal{D}_{100,50}$; Table~\ref{tab:sim_power}), so larger panels allow these effect sizes to be interpreted rather than merely detected. Since both models are estimable within the same routine, the comparison can be carried out directly on any given dataset rather than settled a priori.

\section{Discussion}\label{sec:discuss}
Many psychological change processes are characterized by discontinuous state transitions. The flexibility of RSSS modeling framework allows researchers to include interactions between the intra-individual and inter-individual level. This is often necessary in order to validly represent the dynamic underlying processes (e.g., where baseline covariates modulate how current affective states impact decisions). We presented a frequentist approach to estimating such flexible RSSS models. The development of the estimation procedure enables forecasting and inferences in real-time. 

The extended Kim filter proposed here, serves as an effective tool for detecting unobserved change processes or regime-switches. The effectiveness is particularly utilized when such changes manifest through variations at the within- and between-levels in each individual. The described model and estimation procedures can accommodate a wide variety of dynamic models.

Our model can be readily extended to accommodate three or more regimes, or a more flexible state-space representation, such as incorporating a higher order autoregression, within-between interaction effects, or a polynomial function \citep[cf.,][]{Chow2013}. Another interesting extension is to extend the Markov-switching models: incorporating higher-order autoregressive terms and their interactions, or using polynomial functions, will make the model transition mechanism more flexible. 

The simulation study employed a $2 \times 2$ factorial design crossing sample size $N \in \{50, 100\}$ and estimation length $T_{\text{est}} \in \{25, 50\}$. In broad terms, and consistent with the general asymptotic theory for mixed-effects and state-space models \citep[cf.,][]{Harvey1990}, increasing $N$ is expected chiefly to sharpen inferences about the between-individual parameters, which are identified from variation across persons, while increasing the number of observed time points $T_{\text{est}}$ is expected chiefly to sharpen inferences about the within-individual parameters, which are identified from temporal variation within each person. In the present study, these expected patterns hold for the well-identified parameters under the integrated estimator: recovery of the autoregressive coefficients and of the variance component $\bm{Q}_2$ improves with both design factors, and the baseline transition rate $\gamma_1$ is detected with power $1.00$ throughout. The exceptions are the within-person transition coefficients, which remain weakly identified across all conditions because the design carries little information about them, as discussed below, and the between-person effects, which remain attenuated by the plug-in factor scores irrespective of $N$ and $T_{\text{est}}$. The same design factors govern statistical power (Table~\ref{tab:sim_power}): power to detect the within-level structural and measurement parameters increased with both larger $N$ and longer $T_{\text{est}}$, whereas the between-level and process-noise variance components retained negligible power throughout, for the reasons given in Section~\ref{sec:sim}.

\subsection*{Open questions}
This article did not evaluate the performance under different degree of missingness. Our empirical and simulation study presented two extremes: one with a large proportion of missingness and the other with a complete dataset. The missing proportion in our empirical dataset differed sharply before and after the dropout (55.8\% versus 88.2\% of entries with at least one variable missing; Section~\ref{sec:motiv}), which implies that the data might not be missing at random (MAR) but missing not at random (MNAR).

Our classification result is dependent on the proportion of clusters. Basically, in the case of imbalanced data, as observed in our simulation studies, it would be more difficult to identify the classes which have smaller sample sizes. Furthermore, due to the nature of clustering algorithms, the performance of our method might be sensitive to the similarities of regime distributions. If the regime distributions are very similar (i.e., distribution with large overlapping), the performance is expected to deteriorate. 

Direct comparison with Bayesian counterpart would be an interesting future direction. In order to make a full comparison, all the Bayesian priors need to be substituted by uninformative ones. Bayesian inference with informative prior is a regularized estimation and will not be equivalent to our fully data-driven frequentist framework. Changing all the informative priors to uninformative ones is expected to be time-consuming due to the curse of dimensionality, given that the current Bayesian implementation \citep{Kelava2022} already requires a lot of time to compute. 

The estimator adopted here marginalizes the person-specific random intercept residuals $\bm{\zeta}_{2i}$ within the filter under their model-implied $\mathcal{N}(\bm{0}, \bm{Q}_2)$ distribution (Section~\ref{sec:integrated}): the variance component enters the likelihood directly, and the marginalization is exact conditional on the regime path. A practical consequence is that the filtered mean of the $\bm{\zeta}_{2i}$ block updates sequentially as observations arrive, so the between-person component adapts online, and its uncertainty is propagated into every innovation covariance rather than being fixed at a point estimate.

The remaining plug-in concerns the between-person factor: the estimation uncertainty of $\hat{\bm{\eta}}_{2i}$ (obtained via Bartlett factor scores) is not propagated into the Kim filter, and the simulation shows that its measurement error attenuates the coefficients that load on it ($\bm{B}_{2s}$, $\gamma_2$) and inflates $P_2$, in close quantitative agreement with the classical attenuation formula (Section~\ref{sec:sim}). Extending the marginalization to $\bm{\eta}_{2i}$, in the same way it is here applied to $\bm{\zeta}_{2i}$, would remove this attenuation; the between-person measurement parameters ($\bm{\Lambda}_2$ and $\bm{R}_2$), currently identified only through the plug-in factor scores rather than from the between-person covariance of $\bm{y}_{2i}$, would then be identified from their own likelihood contribution. Short of full marginalization, method-of-moments corrections that explicitly account for the unreliability of factor score estimates, in the tradition of the corrected factor score regressions of \citet{Skrondal2001} and \citet{Croon2002} and their recent unification in the structural-after-measurement framework of \citet{RosseelLoh2024}, offer a lighter route to de-attenuating $\bm{B}_{2s}$ and $\gamma_2$; adapting these corrections to the present filtering context is an attractive intermediate step. A related refinement concerns the transition probabilities, which are currently evaluated at point estimates of the latent predictors (Section~\ref{sec:est}): because the transition logit is linear in $\bm{\eta}_{1i,t-1}$ given $\bm{\eta}_{2i}$, its expectation over the filtered distribution of $\bm{\eta}_{1i,t-1}$ reduces to a one-dimensional Gaussian integral for each regime pair $(s, s')$, and the same integration extends to the Bartlett score using its known error variance. This is the first item on the remaining agenda below.

The simulation results permit a clean separation of what is settled from what remains. The dynamic and variance components are recovered accurately: $\mathrm{diag}(\bm{B}_{3s})$ with biases of at most $0.10$, and the variance component $\bm{Q}_2$ with a small upward bias and valid standard errors (Section~\ref{sec:sim}), so autoregressive persistence and stable person-specific levels are cleanly separated. Regime classification is correspondingly strong: sensitivity reaches $.69$--$.91$ with specificity of $.87$--$.94$ across conditions and windows (Table~\ref{tab:sim_metrics}); the nonlinear RSSS model of \citet{Chow2013}, by comparison, reports sensitivity of $.14$--$.16$ (with a Type-I error rate near $.01$) and attributes the modest detection to the small separation between the regimes. Specificity remains the weaker component over the forecast window for structural reasons: regime-1 observations become scarce under the near-absorbing structure, and a false positive cannot be reversed once made. Relaxing this assumption, either by allowing a small non-zero switch-back probability or by estimating $P_{12}$ freely, is a natural sensitivity analysis, although a substantively meaningful reversal rate would conflict with the theoretical premise that intention formation is rarely reversed \citep{Heublein2014}. Not settled, but now cleanly characterized, is the weak identification of the within-person transition coefficients ($\bm{\gamma}_3$, $\bm{\gamma}_4$): their biases are moderate and coverage is near nominal, but power stays at or below $.40$ (and below $.20$ outside the largest design), because each person contributes no more than a single transition and the within-person predictor variance is limited. The baseline transition rate $\gamma_1$, by contrast, is well identified (power $1.00$, bias at most $+0.21$), and the restricted comparison shows that omitting the weakly identified predictors changes sensitivity and specificity by at most $.01$. The remaining agenda, in order of expected return, is: (i) extending the marginalization to the between-person factor $\bm{\eta}_{2i}$, which would remove the remaining attenuation of $\bm{B}_{2s}$ and $\gamma_2$ and the inflation of $P_2$, and would identify $\bm{\Lambda}_2$ and $\bm{R}_2$ from their own likelihood; (ii) regularization of the transition coefficients, since $\bm{\gamma}_3$ and $\bm{\gamma}_4$ remain weakly identified because the design carries little information about them; and (iii) in the present application, exact marginalization over regime paths: because regime~2 is absorbing and all persons start in regime~1, only $T+1$ regime histories are possible per person, so the collapsing approximation of step~5 could be removed exactly; this route is specific to absorbing designs and does not generalize to unrestricted transition matrices.

A further limitation concerns the transition-model coefficients $\gamma_2$, $\bm{\gamma}_3$, and $\bm{\gamma}_4$, which link the between-person factor $\bm{\eta}_{2i}$, the within-person latent state $\bm{\eta}_{1i,t-1}$, and their interaction to the timing of the regime transition. These coefficients are weakly identified in the present empirical sample. Two features make this explicit. First, across the five independent optimization runs, which attain essentially the same log-likelihood (all within $\Delta < 0.002$), the $\bm{\gamma}_3$ and $\bm{\gamma}_4$ estimates vary by amounts comparable to their own magnitude, so different coefficient values are near-equally consistent with the data. Second, the outer-product-of-gradients standard errors exceed the point estimates in magnitude throughout, and the corresponding confidence intervals include zero by a wide margin. We therefore do not interpret the sign or magnitude of the individual $\bm{\gamma}_3$ and $\bm{\gamma}_4$ estimates and report them only with their (large) standard errors; the baseline transition rate $\gamma_1$ is the one exception, being well identified in the simulation and empirically estimated at $1.53$ (SE $= 0.63$), consistent with the posterior mean of $1.48$ in \citet{Kelava2022}. The residual uncertainty of the within-person transition effects reflects the limited information available for them in this sample rather than a defect of the estimator: the simulation shows the same weak identification under known population values (power $\leq .40$ for $\bm{\gamma}_3$ and $\bm{\gamma}_4$), and, because the transition is one-directional ($P_{12} \approx 0$; Assumption~5), each person contributes no more than a single transition. By contrast, in the Bayesian counterpart \citep{Kelava2022}, the $\mathcal{N}(0,1)$ priors on $\bm{\gamma}_3$ and $\bm{\gamma}_4$ act as implicit regularization, stabilizing the posterior at substantially smaller posterior standard deviations. Achieving comparable precision in the frequentist setting would require a substantially larger or more informative sample; explicit regularization, such as a weakly-informative prior or a ridge-type penalty on the transition-probability coefficients, is one option whose development and evaluation we leave for future work.

Lastly, identifying the correct model is not trivial in practice, as any model is an approximation in social science \citep{browne1993alternative}. Model estimation using a misspecified model is known to harm the prediction in SEM \citep{rooij2023}. In longitudinal models, structural misspecification can be found in many parts including measurement model (e.g., omitting lagged cross-loadings), time-series model (e.g., choosing wrong order in VAR), and model heterogeneity (e.g., ignoring within-person model switch). Under model uncertainty, model stacking approaches \citep[e.g.,][]{yao2021bayesian} may offer an alternative solution by accommodating many ($\gg 2$) models/regimes in a flexible manner.

\begin{Backmatter}

\paragraph{Acknowledgments}
We thank the Editor, Associate Editor, and three anonymous referees for their careful review and valuable comments.
Computations were performed on the bwForCluster Helix at Heidelberg University. The empirical estimation took less than one hour; the simulation study required approximately 8 hours across all 400 replications.
During the preparation of this manuscript, the authors used Claude (claude-sonnet-4-6, claude-opus-4-8, and claude-fable-5, Anthropic; accessed via the Cowork desktop application, May--August 2026) to assist with manuscript editing, response letter preparation, \LaTeX{} code review, and the implementation and post-processing of the estimation and simulation code. All AI-assisted text, code, and results were reviewed, verified, and revised by the authors, who take full responsibility for the final manuscript.

\paragraph{Funding Statement}
The authors acknowledge support by the state of Baden-W\"{u}rttemberg through bwHPC and the German Research Foundation (DFG) through grant INST 35/1597-1 FUGG.

\paragraph{Competing Interests}
The authors have no financial or non-financial competing interests to disclose.

\paragraph{Data Availability Statement}
The associated replication files for the simulation study, including all analysis code, are available online at \url{https://github.com/PsychometricsMZ/Regime-switching}. 
The empirical data used for the application study cannot be made publicly available due to the terms of the data disclosure agreement and privacy concerns.

\printbibliography

\end{Backmatter}

\clearpage
\appendix
\makeatletter\@addtoreset{table}{section}\makeatother
\renewcommand{\thetable}{\thesection\arabic{table}}
\section{Notation Summary}
\label{app:notation}

Table~\ref{tab:notation} provides a summary of the major symbols used throughout the paper, including their dimensions and interpretations.

\begin{table}[H]
    \centering
    \small
    \caption{Summary of notation. Subscripts $i$, $t$, and $s$ index individual ($i=1,\ldots,N$), time point ($t=1,\ldots,T$), and regime ($s \in \{1,2\}$), respectively}
    \label{tab:notation}
    \begin{tabular}{@{}lll@{}}
        \toprule
        \textbf{Symbol} & \textbf{Dimension} & \textbf{Description} \\
        \midrule
        \multicolumn{3}{@{}l}{\textit{Index quantities}} \\
        $N$ & --- & Number of individuals \\
        $T$ & --- & Number of time points \\
        $O_1$ & --- & Number of time-varying observed variables \\
        $U_1$ & --- & Number of dynamic (within-level) latent variables \\
        $O_2$ & --- & Number of time-invariant observed variables \\
        $U_2$ & --- & Number of time-invariant (between-level) latent variables \\
        \midrule
        \multicolumn{3}{@{}l}{\textit{Observed and latent variables}} \\
        $\bm{y}_{1it}$ & $O_1 \times 1$ & Time-varying observed variables for person $i$ at time $t$ \\
        $\bm{y}_{2i}$  & $O_2 \times 1$ & Time-invariant observed variables for person $i$ \\
        $\bm{\eta}_{1it}$ & $U_1 \times 1$ & Dynamic latent variables (within-level) for person $i$ at time $t$ \\
        $\bm{\eta}_{2i}$  & $U_2 \times 1$ & Time-invariant latent variables (between-level) for person $i$ \\
        $S_{it}$ & scalar & Discrete latent regime for person $i$ at time $t$; $S_{it} \in \{1,2\}$ \\
        \midrule
        \multicolumn{3}{@{}l}{\textit{Measurement model parameters}} \\
        $\bm{\Lambda}_{1s}$ & $O_1 \times U_1$ & Factor loading matrix for within-level model (regime $s$) \\
        $\bm{R}_{1s}$       & $O_1 \times O_1$ & Residual covariance matrix for within-level model (regime $s$) \\
        $\bm{\Lambda}_{2}$  & $O_2 \times U_2$ & Factor loading matrix for between-level model \\
        $\bm{P}_{2}$        & $U_2 \times U_2$ & Latent factor covariance matrix (between-level) \\
        $\bm{R}_{2}$        & $O_2 \times O_2$ & Residual covariance matrix for between-level model \\
        \midrule
        \multicolumn{3}{@{}l}{\textit{Structural model parameters}} \\
        $\bm{b}_{1s}$    & $U_1 \times 1$      & Conditional intercept vector (regime $s$) \\
        $\bm{B}_{2s}$    & $U_1 \times U_2$    & Effect of between-level factors on intercepts (regime $s$) \\
        $\bm{\zeta}_{2i}$ & $U_1 \times 1$     & Person-specific random intercept residual \\
        $\bm{Q}_{2}$     & $U_1 \times U_1$    & Covariance matrix of random intercept residuals \\
        $\bm{B}_{3s}$    & $U_1 \times U_1$    & Mean VAR(1) coefficient matrix (regime $s$) \\
        $\bm{B}_{4s}$    & $U_1 \times U_1 \times U_2$ & Interaction effect tensor: moderation of VAR(1) by $\bm{\eta}_{2i}$ (regime $s$) \\
        $\bm{Q}_{1s}$    & $U_1 \times U_1$    & Process noise covariance matrix (regime $s$) \\
        \midrule
        \multicolumn{3}{@{}l}{\textit{Markov-switching model parameters}} \\
        $\gamma_{s'1}$      & scalar           & Intercept in transition probability from regime $s'$ \\
        $\bm{\gamma}_{s'2}$      & $1 \times U_2$   & Effect of between-level factors on transition probability \\
        $\bm{\gamma}_{s'3}$ & $1 \times U_1$   & Effect of within-level factors on transition probability \\
        $\bm{\gamma}_{s'4}$ & $1 \times U_1$   & Interaction effects on transition probability \\
        $\sigma(\cdot)$     & ---              & Sigmoid function: $\sigma(x) = [1 + \exp(-x)]^{-1}$ \\
        \bottomrule
    \end{tabular}
\end{table}

\section{Parameter Estimates: Empirical Study}
\label{app:empirical}

\emph{Optimization details.} Parameter optimization used Rprop \citep{Riedmiller1993} with hyperparameters $\left( \Delta_{0}=0.005, \bm{\eta} = (\eta_{+},\eta_{-}) = (1.2, 0.5), (\Delta_{\text{min}}, \Delta_{\text{max}})=(10^{-6}, 50) \right)$. If $\Delta \log f(\bm{Y}_1 | \Theta_{k}) < 10^{-4}$ continues 20 steps in a row, the optimization is terminated (see Equation \ref{conv}). To keep the model-implied initialization finite when the optimizer transiently explores the boundary of the stationary region, the autoregressive diagonal entering $(\bm{I} - \bm{B}_{3i1})^{-1}$ and the stationary covariance is clipped to $|(\bm{B}_{3i1})_{jj}| \leq 0.95$; at the estimated values this clip is at most negligibly active and does not affect the reported results.

\begin{table}
    \centering
    \caption{Parameter estimates for the Markov-switching model (best of five runs). Parameters with asterisk are fixed; parameters marked $\dagger$ are fixed to zero for the motivational variables (see Assumption~6). The transition coefficients other than $\gamma_1$ are weakly identified: standard errors exceed the estimates, and the estimates vary across near-equivalent optima (Section~\ref{sec:emp:impl})}
    \label{tab:emp_params_ms}
    \begin{tabular}{@{}llll@{}}
        \toprule
        \textbf{Parameter} & \textbf{Index} & \textbf{Estimates} & \textbf{SE} \\
        \midrule
        $\gamma_1$
         & 1 & 1.53 & 0.63 \\
        $\gamma_2$
         & 1 & $-0.11$ & 1.51 \\
        $\bm{\gamma}_3$
         & 1 & $-0.52$ & 3.75 \\
         & 2 & $-1.67$ & 1.79 \\
         & 3 & $-1.19$ & 2.46 \\
         & 4 & 0.25 & 1.74 \\
         & 5 & 0.96 & 2.08 \\
         & 6 & 0.06 & 1.41 \\
         & 7 & $-0.20$ & 1.42 \\
        $\bm{\gamma}_4$
         & 1 & $0^\dagger$ & --- \\
         & 2 & 0.08 & 2.51 \\
         & 3 & 0.61 & 3.79 \\
         & 4 & $0^\dagger$ & --- \\
         & 5 & $-0.85$ & 3.44 \\
         & 6 & $0^\dagger$ & --- \\
         & 7 & $0^\dagger$ & --- \\
        $P_{12}$
         & 1 & $10^{-12}$* & --- \\
        \bottomrule
    \end{tabular}
\end{table}

\begin{table}
    \centering
    \caption{Parameter estimates for the structural parameters ($s=1$). The autoregressive coefficient of the third within-scale (\textit{afraid to fail}) is not identified because its process-noise variance is degenerate (Section~\ref{sec:emp:sm})}
    \label{tab:emp_params_sm1}
    \begin{tabular}{@{}llll@{}}
        \toprule
        \textbf{Parameter} & \textbf{Index} & \textbf{Estimates} & \textbf{SE} \\
        \midrule
        $\bm{b}_{11}$
         & 1 & 0.03 & 0.04 \\
         & 2 & $-0.05$ & 0.21 \\
         & 3 & $-0.23$ & 0.22 \\
         & 4 & $-0.05$ & 0.14 \\
         & 5 & $-0.04$ & 0.14 \\
         & 6 & $-0.02$ & 0.13 \\
         & 7 & $-0.21$ & 0.25 \\
        $\bm{b}_{21}$
         & 1 & $-0.00$ & 0.04 \\
         & 2 & $-0.16$ & 0.23 \\
         & 3 & $-0.26$ & 0.44 \\
         & 4 & $-0.11$ & 0.16 \\
         & 5 & $-0.12$ & 0.36 \\
         & 6 & $-0.02$ & 0.21 \\
         & 7 & $-0.16$ & 0.39 \\
        $\mathrm{diag}(\bm{B}_{31})$
         & 1 & 0.77 & 0.17 \\
         & 2 & 0.32 & 0.36 \\
         & 3 & $-0.12$ & 0.50 \\
         & 4 & 0.47 & 0.37 \\
         & 5 & 0.24 & 0.34 \\
         & 6 & 0.26 & 0.37 \\
         & 7 & 0.23 & 0.48 \\
        $\mathrm{diag}(\bm{B}_{41})$
         & 1 & 0.04 & 0.19 \\
         & 2 & 0.01 & 0.19 \\
         & 3 & $-0.08$ & 0.59 \\
         & 4 & $-0.08$ & 0.33 \\
         & 5 & $-0.09$ & 0.44 \\
         & 6 & 0.04 & 0.42 \\
         & 7 & $-0.05$ & 0.61 \\
        \bottomrule
    \end{tabular}
\end{table}

\begin{table}
    \centering
    \caption{Parameter estimates for the structural parameters ($s=2$). $\bm{b}_{12}$ is derived as $\bm{b}_{11} + \exp(\delta)$ and has no SE}
    \label{tab:emp_params_sm2}
    \begin{tabular}{@{}llll@{}}
        \toprule
        \textbf{Parameter} & \textbf{Index} & \textbf{Estimates} & \textbf{SE} \\
        \midrule
        $\bm{b}_{12}$
         & 1 & 0.08 & --- \\
         & 2 & 0.21 & --- \\
         & 3 & 0.28 & --- \\
         & 4 & 0.24 & --- \\
         & 5 & 0.44 & --- \\
         & 6 & 0.27 & --- \\
         & 7 & 0.08 & --- \\
        $\bm{b}_{22}$
         & 1 & $-0.02$ & 0.07 \\
         & 2 & $-0.15$ & 0.31 \\
         & 3 & $-0.23$ & 0.32 \\
         & 4 & $-0.14$ & 0.23 \\
         & 5 & $-0.15$ & 0.33 \\
         & 6 & 0.02 & 0.22 \\
         & 7 & $-0.12$ & 0.32 \\
        $\mathrm{diag}(\bm{B}_{32})$
         & 1 & 0.84 & 0.12 \\
         & 2 & 0.49 & 0.25 \\
         & 3 & 0.43 & 0.18 \\
         & 4 & 0.46 & 0.35 \\
         & 5 & 0.44 & 0.15 \\
         & 6 & 0.49 & 0.20 \\
         & 7 & 0.47 & 0.23 \\
        $\mathrm{diag}(\bm{B}_{42})$
         & 1 & 0.01 & 0.13 \\
         & 2 & 0.08 & 0.22 \\
         & 3 & 0.05 & 0.20 \\
         & 4 & $-0.00$ & 0.22 \\
         & 5 & 0.04 & 0.31 \\
         & 6 & 0.06 & 0.26 \\
         & 7 & 0.07 & 0.14 \\
        \bottomrule
    \end{tabular}
\end{table}

\begin{table}
    \centering
    \caption{Parameter estimates for the process noise variances of the structural model. $\bm{Q}_2$ is estimated by maximum likelihood within the filter, and its standard errors are obtained from the same OPG computation as the remaining parameters (Section~\ref{sec:integrated})}
    \label{tab:emp_params_sm3}
    \begin{tabular}{@{}llll@{}}
        \toprule
        \textbf{Parameter} & \textbf{Index} & \textbf{Estimates} & \textbf{SE} \\
        \midrule
        $\bm{Q}_1$
         & 1 & 0.021 & 0.50 \\
         & 2 & 0.078 & 0.39 \\
         & 3 & 0.000 & 0.00 \\
         & 4 & 0.032 & 0.72 \\
         & 5 & 0.058 & 0.37 \\
         & 6 & 0.163 & 0.41 \\
         & 7 & 0.173 & 0.16 \\
        $\bm{Q}_2$
         & 1 & 0.003 & 1.82 \\
         & 2 & 0.104 & 0.78 \\
         & 3 & 0.152 & 0.90 \\
         & 4 & 0.044 & 1.60 \\
         & 5 & 0.116 & 0.82 \\
         & 6 & 0.109 & 0.87 \\
         & 7 & 0.194 & 1.07 \\
        \bottomrule
    \end{tabular}
\end{table}

\begin{table}
    \centering
    \caption{Parameter estimates for the factor loadings}
    \label{tab:emp_params_mm1}
    \begin{tabular}{@{}llll@{}}
        \toprule
        \textbf{Parameter} & \textbf{Index} & \textbf{Estimates} & \textbf{SE} \\
        \midrule
        $\bm{\Lambda}_{1, \text{free}}$
         & 1 & 1.43 & 0.11 \\
         & 2 & 1.11 & 0.11 \\
         & 3 & 0.90 & 0.12 \\
         & 4 & 0.91 & 0.09 \\
         & 5 & 1.18 & 0.08 \\
         & 6 & 1.10 & 0.08 \\
        \bottomrule
    \end{tabular}
\end{table}

\begin{table}
    \centering
    \caption{Parameter estimates for the residual variances of the measurement model and the between-level factor variance. Note: $\bm{R}_2$ standard errors are very large, because $\bm{\Lambda}_2$ and $\bm{R}_2$ remain identified only through the plug-in factor scores (Section~\ref{sec:integrated})}
    \label{tab:emp_params_mm2}
    \begin{tabular}{@{}llll@{}}
        \toprule
        \textbf{Parameter} & \textbf{Index} & \textbf{Estimates} & \textbf{SE} \\
        \midrule
        $\bm{R}_1$
         & 1 & 0.38 & 0.14 \\
         & 2 & 0.26 & 0.13 \\
         & 3 & 0.52 & 0.10 \\
         & 4 & 0.24 & 0.18 \\
         & 5 & 0.25 & 0.13 \\
         & 6 & 0.32 & 0.10 \\
         & 7 & 0.39 & 0.10 \\
         & 8 & 0.31 & 0.17 \\
         & 9 & 0.36 & 0.19 \\
         & 10 & 0.29 & 0.18 \\
         & 11 & 0.22 & 0.20 \\
         & 12 & 0.33 & 0.27 \\
         & 13 & 0.51 & 0.17 \\
         & 14 & 0.66 & 0.05 \\
         & 15 & 0.55 & 0.11 \\
         & 16 & 0.23 & 0.21 \\
         & 17 & 0.48 & 0.11 \\
        $\bm{R}_2$
         & 1 & 1.81 & 285.92 \\
         & 2 & 0.32 & 285.92 \\
        $\bm{P}_2$
         & 1 & 0.87 & --- \\
        \bottomrule
    \end{tabular}
\end{table}

\section{Parameter Estimates: Simulation Studies}
\label{app:simulation}

\begin{table}
    \centering
    \footnotesize
    \caption{Summary statistics for Markov-switching model parameter estimates across the four simulation conditions under the integrated estimator ($R=100$ replications per condition, of which 99 converged; $\gamma_1$ is freely estimated). $P_{12}^{*}$ is fixed for identification. Coverage rate is the empirical proportion of 95\% confidence intervals that contain the true value (nominal: 0.95)}
    \label{tab:sim_params_ms}
    \resizebox{\textwidth}{!}{%
    \begin{tabular}{@{}lll rrr rrr rrr rrr@{}}
        \toprule
        \textbf{Parameter} & \textbf{Idx} & \textbf{True $\theta$} &
        \multicolumn{3}{c}{$\mathcal{D}_{50,25}$} &
        \multicolumn{3}{c}{$\mathcal{D}_{100,25}$} &
        \multicolumn{3}{c}{$\mathcal{D}_{50,50}$} &
        \multicolumn{3}{c}{$\mathcal{D}_{100,50}$} \\
        \cmidrule(lr){4-6}\cmidrule(lr){7-9}\cmidrule(lr){10-12}\cmidrule(lr){13-15}
        & & & Bias & RMSE & Cov. & Bias & RMSE & Cov. & Bias & RMSE & Cov. & Bias & RMSE & Cov. \\
        \midrule
        $\gamma_1$
         & 1 & $3.50$ & $0.20$ & 0.54 & 0.97 & $0.08$ & 0.26 & 0.93 & $0.17$ & 0.38 & 0.95 & $0.21$ & 0.27 & 0.79 \\
        $\gamma_2$
         & 1 & $-1.17$ & $1.02$ & 1.25 & 0.78 & $1.05$ & 1.15 & 0.20 & $1.02$ & 1.15 & 0.51 & $1.03$ & 1.06 & 0.03 \\
        $\bm{\gamma}_3$
         & 1 & $0.26$ & $0.13$ & 0.70 & 0.99 & $0.04$ & 0.43 & 0.94 & $0.13$ & 0.46 & 0.99 & $0.13$ & 0.34 & 0.87 \\
         & 2 & $-0.83$ & $0.32$ & 0.93 & 0.99 & $0.46$ & 0.63 & 0.87 & $0.35$ & 0.69 & 0.97 & $0.47$ & 0.59 & 0.73 \\
        $\bm{\gamma}_4$
         & 1 & $-0.74$ & $0.49$ & 1.20 & 0.98 & $0.59$ & 0.84 & 0.82 & $0.52$ & 0.87 & 0.95 & $0.48$ & 0.61 & 0.75 \\
         & 2 & $-0.19$ & $0.21$ & 1.37 & 0.97 & $0.03$ & 0.66 & 0.96 & $0.16$ & 0.83 & 0.99 & $0.15$ & 0.48 & 0.95 \\
        $P_{12}^{*}$
         & 1 & $0.00$ & $-0.00$ & 0.00 & --- & $-0.00$ & 0.00 & --- & $-0.00$ & 0.00 & --- & $-0.00$ & 0.00 & --- \\
        \bottomrule
    \end{tabular}
    }%
\end{table}

\begin{table}
    \centering
    \footnotesize
    \caption{Summary statistics for structural model parameter estimates across the four simulation conditions under the integrated estimator. Coverage rate is the empirical proportion of 95\% confidence intervals containing the true value (nominal: 0.95); entries marked (---) indicate parameters for which standard errors were not computed. The residual biases of $\bm{b}_{21}$, $\bm{b}_{22}$, and $P_2$ are the plug-in measurement error of the Bartlett factor scores (Section~\ref{sec:sim})}
    \label{tab:sim_params_sm}
    \resizebox{\textwidth}{!}{%
    \begin{tabular}{@{}lll rrr rrr rrr rrr@{}}
        \toprule
        \textbf{Parameter} & \textbf{Idx} & \textbf{True $\theta$} &
        \multicolumn{3}{c}{$\mathcal{D}_{50,25}$} &
        \multicolumn{3}{c}{$\mathcal{D}_{100,25}$} &
        \multicolumn{3}{c}{$\mathcal{D}_{50,50}$} &
        \multicolumn{3}{c}{$\mathcal{D}_{100,50}$} \\
        \cmidrule(lr){4-6}\cmidrule(lr){7-9}\cmidrule(lr){10-12}\cmidrule(lr){13-15}
        & & & Bias & RMSE & Cov. & Bias & RMSE & Cov. & Bias & RMSE & Cov. & Bias & RMSE & Cov. \\
        \midrule
        $\bm{b}_{11}$
         & 1 & $-0.05$ & $-0.00$ & 0.07 & 1.00 & $-0.01$ & 0.05 & 0.98 & $-0.01$ & 0.07 & 0.99 & $-0.00$ & 0.06 & 0.98 \\
         & 2 & $-0.06$ & $0.03$ & 0.07 & 1.00 & $0.03$ & 0.05 & 0.96 & $0.03$ & 0.07 & 0.99 & $0.03$ & 0.06 & 0.94 \\
        $\bm{b}_{12}$
         & 1 & $0.45$ & $-0.07$ & 0.14 & --- & $-0.08$ & 0.12 & --- & $-0.07$ & 0.12 & --- & $-0.07$ & 0.11 & --- \\
         & 2 & $0.22$ & $0.05$ & 0.12 & --- & $0.05$ & 0.08 & --- & $0.03$ & 0.08 & --- & $0.03$ & 0.06 & --- \\
        $\bm{b}_{21}$
         & 1 & $-0.87$ & $0.62$ & 0.63 & 0.06 & $0.61$ & 0.62 & 0.00 & $0.61$ & 0.62 & 0.09 & $0.61$ & 0.61 & 0.00 \\
         & 2 & $-0.77$ & $0.54$ & 0.55 & 0.11 & $0.54$ & 0.54 & 0.00 & $0.54$ & 0.55 & 0.04 & $0.54$ & 0.55 & 0.00 \\
        $\bm{b}_{22}$
         & 1 & $-0.92$ & $0.59$ & 0.61 & 0.39 & $0.59$ & 0.60 & 0.02 & $0.58$ & 0.60 & 0.10 & $0.57$ & 0.58 & 0.00 \\
         & 2 & $-0.45$ & $0.27$ & 0.29 & 0.85 & $0.28$ & 0.29 & 0.21 & $0.28$ & 0.29 & 0.64 & $0.29$ & 0.29 & 0.03 \\
        $\mathrm{diag}(\bm{B}_{31})$
         & 1 & $0.33$ & $0.01$ & 0.10 & 0.98 & $0.03$ & 0.07 & 0.89 & $0.02$ & 0.08 & 0.93 & $0.04$ & 0.07 & 0.73 \\
         & 2 & $0.15$ & $0.02$ & 0.13 & 0.97 & $0.03$ & 0.09 & 0.93 & $0.06$ & 0.11 & 0.92 & $0.08$ & 0.11 & 0.66 \\
        $\mathrm{diag}(\bm{B}_{32})$
         & 1 & $0.31$ & $0.08$ & 0.16 & 0.90 & $0.10$ & 0.14 & 0.56 & $0.09$ & 0.14 & 0.63 & $0.10$ & 0.14 & 0.40 \\
         & 2 & $0.69$ & $-0.02$ & 0.06 & 0.99 & $-0.01$ & 0.05 & 0.96 & $0.00$ & 0.04 & 0.96 & $0.01$ & 0.02 & 0.93 \\
        $\mathrm{diag}(\bm{B}_{41})$
         & 1 & $0.11$ & $-0.06$ & 0.11 & 0.98 & $-0.07$ & 0.09 & 0.82 & $-0.06$ & 0.09 & 0.96 & $-0.04$ & 0.08 & 0.64 \\
         & 2 & $-0.30$ & $0.22$ & 0.25 & 0.79 & $0.23$ & 0.24 & 0.17 & $0.22$ & 0.25 & 0.62 & $0.23$ & 0.24 & 0.05 \\
        $\mathrm{diag}(\bm{B}_{42})$
         & 1 & $0.57$ & $-0.35$ & 0.38 & 0.51 & $-0.34$ & 0.35 & 0.02 & $-0.33$ & 0.35 & 0.06 & $-0.31$ & 0.32 & 0.02 \\
         & 2 & $0.01$ & $0.00$ & 0.06 & 0.99 & $-0.00$ & 0.04 & 0.96 & $0.00$ & 0.04 & 0.97 & $-0.00$ & 0.03 & 0.89 \\
        $\bm{Q}_1$
         & 1 & $0.08$ & $0.01$ & 0.01 & 1.00 & $0.01$ & 0.01 & 1.00 & $0.00$ & 0.01 & 1.00 & $0.01$ & 0.01 & 1.00 \\
         & 2 & $0.05$ & $0.00$ & 0.01 & 1.00 & $0.00$ & 0.01 & 1.00 & $0.00$ & 0.01 & 1.00 & $0.00$ & 0.01 & 1.00 \\
        $\bm{Q}_2$
         & 1 & $0.16$ & $0.03$ & 0.07 & 1.00 & $0.04$ & 0.06 & 1.00 & $0.03$ & 0.06 & 1.00 & $0.05$ & 0.07 & 1.00 \\
         & 2 & $0.12$ & $0.04$ & 0.07 & 1.00 & $0.04$ & 0.06 & 1.00 & $0.03$ & 0.05 & 1.00 & $0.02$ & 0.04 & 1.00 \\
        $P_2$
         & 1 & $0.20$ & $0.51$ & 0.54 & --- & $0.46$ & 0.47 & --- & $0.48$ & 0.52 & --- & $0.46$ & 0.48 & --- \\
        \bottomrule
    \end{tabular}
    }%
\end{table}

\begin{table}
    \centering
    \footnotesize
    \caption{Summary statistics for measurement model parameter estimates across the four simulation conditions under the integrated estimator. Coverage rates for $\bm{R}_1$ are at 1.00 (conservative intervals), whereas the between-level residual variances $\bm{R}_2$ are weakly identified}
    \label{tab:sim_params_mm}
    \resizebox{\textwidth}{!}{%
    \begin{tabular}{@{}lll rrr rrr rrr rrr@{}}
        \toprule
        \textbf{Parameter} & \textbf{Idx} & \textbf{True $\theta$} &
        \multicolumn{3}{c}{$\mathcal{D}_{50,25}$} &
        \multicolumn{3}{c}{$\mathcal{D}_{100,25}$} &
        \multicolumn{3}{c}{$\mathcal{D}_{50,50}$} &
        \multicolumn{3}{c}{$\mathcal{D}_{100,50}$} \\
        \cmidrule(lr){4-6}\cmidrule(lr){7-9}\cmidrule(lr){10-12}\cmidrule(lr){13-15}
        & & & Bias & RMSE & Cov. & Bias & RMSE & Cov. & Bias & RMSE & Cov. & Bias & RMSE & Cov. \\
        \midrule
        $\bm{R}_1$
         & 1 & $0.23$ & $0.00$ & 0.01 & 1.00 & $0.00$ & 0.01 & 1.00 & $0.00$ & 0.01 & 1.00 & $0.00$ & 0.01 & 1.00 \\
         & 2 & $0.21$ & $-0.00$ & 0.01 & 1.00 & $0.00$ & 0.01 & 1.00 & $0.00$ & 0.01 & 1.00 & $0.00$ & 0.01 & 1.00 \\
         & 3 & $0.16$ & $0.00$ & 0.01 & 1.00 & $0.00$ & 0.01 & 1.00 & $0.00$ & 0.01 & 1.00 & $0.00$ & 0.01 & 1.00 \\
         & 4 & $0.61$ & $0.00$ & 0.02 & 1.00 & $-0.00$ & 0.02 & 1.00 & $0.00$ & 0.02 & 1.00 & $-0.00$ & 0.01 & 1.00 \\
        $\bm{R}_2$
         & 1 & $0.82$ & $0.75$ & 2.69 & 1.00 & $0.34$ & 1.50 & 1.00 & $0.52$ & 2.26 & 1.00 & $0.26$ & 1.19 & 1.00 \\
         & 2 & $0.73$ & $0.26$ & 1.44 & 1.00 & $-0.03$ & 0.29 & 1.00 & $0.25$ & 1.64 & 1.00 & $0.01$ & 0.31 & 1.00 \\
        \bottomrule
    \end{tabular}
    }%
\end{table}

\begin{table}
    \centering
    \footnotesize
    \caption{Power for the simulation parameters across the four conditions under the integrated estimator. Power is the proportion of the (converged) replications in which the 95\% confidence interval excludes zero (i.e., the parameter is detected as nonzero). $P_{12}$ (fixed for identification) and the parameters not included in the power computation ($\bm{b}_{12}$, $\bm{Q}_{2}$, $P_{2}$) are omitted}
    \label{tab:sim_power}
    \begin{tabular}{@{}ll cccc@{}}
        \toprule
        \textbf{Parameter} & \textbf{True $\theta$} & $\mathcal{D}_{50,25}$ & $\mathcal{D}_{100,25}$ & $\mathcal{D}_{50,50}$ & $\mathcal{D}_{100,50}$ \\
        \midrule
        \multicolumn{6}{@{}l}{\textit{Markov-switching parameters}} \\
        $\gamma_{1}$ & $3.50$ & 1.00 & 1.00 & 1.00 & 1.00 \\
        $\gamma_{2}$ & $-1.17$ & 0.01 & 0.08 & 0.02 & 0.10 \\
        $\bm{\gamma}_{3,1}$ & $0.26$ & 0.03 & 0.10 & 0.03 & 0.40 \\
        $\bm{\gamma}_{3,2}$ & $-0.83$ & 0.01 & 0.05 & 0.07 & 0.19 \\
        $\bm{\gamma}_{4,1}$ & $-0.74$ & 0.01 & 0.12 & 0.01 & 0.13 \\
        $\bm{\gamma}_{4,2}$ & $-0.19$ & 0.03 & 0.05 & 0.00 & 0.03 \\
        \midrule
        \multicolumn{6}{@{}l}{\textit{Structural model parameters}} \\
        $\bm{b}_{11,1}$ & $-0.05$ & 0.01 & 0.08 & 0.02 & 0.11 \\
        $\bm{b}_{11,2}$ & $-0.06$ & 0.01 & 0.03 & 0.02 & 0.07 \\
        $\bm{b}_{21,1}$ & $-0.87$ & 0.13 & 0.82 & 0.20 & 0.84 \\
        $\bm{b}_{21,2}$ & $-0.77$ & 0.12 & 0.87 & 0.20 & 0.91 \\
        $\bm{b}_{22,1}$ & $-0.92$ & 0.20 & 0.88 & 0.43 & 0.97 \\
        $\bm{b}_{22,2}$ & $-0.45$ & 0.00 & 0.35 & 0.03 & 0.58 \\
        $\mathrm{diag}(\bm{B}_{31})_1$ & $0.33$ & 0.85 & 1.00 & 0.99 & 1.00 \\
        $\mathrm{diag}(\bm{B}_{31})_2$ & $0.15$ & 0.10 & 0.55 & 0.32 & 0.96 \\
        $\mathrm{diag}(\bm{B}_{32})_1$ & $0.31$ & 0.66 & 1.00 & 1.00 & 1.00 \\
        $\mathrm{diag}(\bm{B}_{32})_2$ & $0.69$ & 0.99 & 1.00 & 1.00 & 1.00 \\
        $\mathrm{diag}(\bm{B}_{41})_1$ & $0.11$ & 0.02 & 0.13 & 0.10 & 0.46 \\
        $\mathrm{diag}(\bm{B}_{41})_2$ & $-0.30$ & 0.03 & 0.21 & 0.09 & 0.37 \\
        $\mathrm{diag}(\bm{B}_{42})_1$ & $0.57$ & 0.28 & 0.80 & 0.76 & 1.00 \\
        $\mathrm{diag}(\bm{B}_{42})_2$ & $0.01$ & 0.01 & 0.04 & 0.03 & 0.11 \\
        $\bm{Q}_{1,1}$ & $0.08$ & 0.00 & 0.00 & 0.00 & 0.00 \\
        $\bm{Q}_{1,2}$ & $0.05$ & 0.00 & 0.00 & 0.00 & 0.00 \\
        \midrule
        \multicolumn{6}{@{}l}{\textit{Measurement model parameters}} \\
        $\bm{R}_{1,1}$ & $0.23$ & 0.93 & 1.00 & 1.00 & 1.00 \\
        $\bm{R}_{1,2}$ & $0.21$ & 0.75 & 1.00 & 0.98 & 1.00 \\
        $\bm{R}_{1,3}$ & $0.16$ & 0.01 & 0.99 & 0.58 & 1.00 \\
        $\bm{R}_{1,4}$ & $0.61$ & 1.00 & 1.00 & 1.00 & 1.00 \\
        $\bm{R}_{2,1}$ & $0.82$ & 0.00 & 0.00 & 0.00 & 0.01 \\
        $\bm{R}_{2,2}$ & $0.73$ & 0.00 & 0.00 & 0.00 & 0.01 \\
        \bottomrule
    \end{tabular}
\end{table}

\end{document}